\documentclass[entropy,article,accept,pdftex,moreauthors]{Definitions/mdpi} 

\firstpage{1} 
\pubvolume{1}
\issuenum{1}
\articlenumber{0}
\pubyear{2023}
\copyrightyear{2023}
\datereceived{ } 
\daterevised{ }
\dateaccepted{ } 
\datepublished{ } 
\hreflink{https://doi.org/}

\Title{What is mature and what is still emerging in the cryptocurrency market?}

\TitleCitation{What is mature and what still emerging in the cryptocurrency market?}

\Author{Stanisław Drożdż $^{1,2\dagger,\ddagger}$\orcidA{}, Jarosław Kwapień $^{2,\ddagger}$\orcidB{} and Marcin Wątorek $^{1,\ddagger}$*\orcidC{}}

\AuthorNames{Firstname Lastname, Firstname Lastname and Firstname Lastname}

\AuthorCitation{Drożdż, S.; Kwapień J.; Wątorek F.}

\address{%
$^{1}$ \quad Faculty of Computer Science and Telecommunications, Cracow University of Technology, ul. Warszawska 24, 31-155 Kraków, Poland\\
$^{2}$ \quad Complex Systems Theory Department, Institute of Nuclear Physics, Polish Academy of Sciences, ul. Radzikowskiego 152, 31-342 Kraków, Poland}

\corres{Correspondence: stanislaw.drozdz@ifj.edu.pl, jaroslaw.kwapien@ifj.edu.pl, marcin.watorek@pk.edu.pl}

\secondnote{These authors contributed equally to this work.}

\abstract{In relation to the traditional financial markets the cryptocurrency market is a recent invention and the trading dynamics of all its components is readily recorded and stored. This fact opens a unique opportunity to follow the multidimensional trajectory of its development since inception up to the present time. Several main characteristics commonly recognized as financial stylized facts of mature markets are here quantitatively studied. In particular, it is shown that the return distributions, volatility clustering effects and even the temporal multifractal correlations for a few highest capitalization cryptocurrencies largely follow those of the well-established financial markets. The smaller ones are somewhat deficient in this regard, however. They are also not as highly cross-correlated among themselves and with other financial markets as the large ones. Quite generally, volume $V$ impact on price changes $R$ appears much stronger on the cryptocurrency market than in the mature stock markets and scales as $R(V) \sim V^{\alpha}$ with $\alpha \gtrsim 1$.}


\begin{document}

\section{Introduction}

Studying the world cryptocurrency market is welcome for many reasons. Up to now, it constitutes the most spectacular and influential application of the distributed ledger technology called the blockchain, which in the underlying peer-to-peer network allows the same access to information for all participants~\cite{wattenhofer2016science,lantz2020mastering}. Research on blockchain technology is also unique because all related data is publicly available in the form of a history of every operation performed on the network. Furthermore, the tick-by-tick data for each transaction made on the cryptocurrency exchange are freely available using the application programming interfaces (APIs) of a given exchange.

As far as the financial, economic, and, in general terms, the social aspects of cryptocurrencies are concerned, a basic related question that arises is whether such digital products can be considered a commonly accepted means of exchange~\cite{Corbet2019,GilCorderoE-2020a,CachanoskyN-2020a}. This is a complex issue involving many social, economical, and technological factors like trust, perceived risk, peer opinions, transaction security, network size effect, supply elasticity, and so on. But also from a dynamical perspective, for this to apply, a certain level of maturity expressed in terms of market efficiency, liquidity, stability, size, and other characteristics is required~\cite{watorek2021,james2022}. Moreover, the developed markets show several statistical properties that newly-established emerging markets often lack. Among such properties, one can list the so-called financial stylized facts: heavy tails of the probability distribution functions of fixed-time returns, long-term memory of volatility, a hierarchical structure of the asset cross-correlations, multifractality, and a stable (or meta-stable) price impact function~\cite{Ausloos2000,ContR-2001a,LeBaronB-2006a,MoroneA-2008a}.

    There is a growing quantitative evidence that the cryptocurrency market continuously advances on a route to maturity understood as sharing its statistical properties with the traditional financial markets. For instance, the most popular and the oldest cryptocurrency, bitcoin (BTC), has passed through two stages of shaping of its probability distribution function (pdf). It started as an extremely volatile asset with the pdf tails that used to decline according to a power law with the exponent reaching almost the L\'evy-stable regime (the L\'evy parameter $\alpha \approx 2$) on short time scales over the years 2012-2013, but then already in the years 2014-2015 the tails of its pdf became thinner and reached the inverse cubic behaviour that is observed universally on the traditional financial markets~\cite{Watorek2021distr}. From that moment on, BTC has maintained this property over the subsequent years~\cite{BEGUSIC2018,DrozdzBTC2018,watorek2021}. The difference between BTC and the traditional assets is that the inverse cubic behaviour of the BTC pdf tails was reported to be preserved up to much longer sampling intervals due to its less frequent trading~\cite{Watorek2021distr}. Similar effects were seen for other major crypto currencies such as ETH~\cite{Watorek2021distr,pessa2023age}. Since BTC and the other cryptocurrencies are traded on many independent platforms that differ in trading frequency, the pdf properties of the same cryptocurrency can be different on different platforms~\cite{watorek2021}. This is quite a unique trait of the cryptocurrencies not observed, for example, in the stock markets and Forex. Heavy pdf tails were also found in time series of volume traded in time unit~\cite{GopikrishnanP-2000a,Plerou2004}, even in the case of cryptocurrencies~\cite{KwapienJ-2022a,Navarro2023}. These two quantities: the log-returns and volume are related with each other, because the size of a trade can have a profound impact on price variation: large trades lead to large price jumps on average (although this relation might be more subtle~\cite{Gillemot2006,BouchaudJP-2010a,TothB-2011a}). Some authors argue that price impact assumes a functional form with square-root dependence of the log-returns on volume~\cite{Gabaix2003,RakR-2013a,Bucci2019} but the others are cautious~\cite{BouchaudJP-2010a,TothB-2011a,Zarinell2015}. 

The long-term memory of volatility fluctuations is responsible for the effect of volatility clustering, i.e., periods of volatile market with large-amplitude fluctuations are interwoven with periods of relatively tranquil dynamics. In addition, the volatility autocorrelation is of a power-law form~\cite{Gopikrishnan1999}. This property has been seen in all financial markets and has also been found in the cryptocurrency dynamics~\cite{DrozdzBTC2018}. The range of memory is comparable in this case with the range for the stock and Forex markets~\cite{DrozdzS-2010a,DrozdzS-2019a}. The scale-free form of the autocorrelation function is connected to fractality, which also requires long-term or long-range correlations to be self-similar. The log-return fluctuations for all the traditional financial markets studied so far show multiscaling together with some other quantities like the inter-transaction times~\cite{MatiaK-2003a,Kwapien2005,oswiecimka2005}. Consistently, multifractal properties have been observed in the cryptocurrency market returns and inter-transaction times for different assets~\cite{takaishi2018statistical,kristj2019,han2020long,takaishi2020market,BARIVIERA2021,Takaishi2021,watorek2021,KwapienJ-2022a,KAKINAKA2022}. Apart from the univariate multiscaling, its bivariate version has also been reported between the log-returns for different cryptocurrencies: BTC and ETH~\cite{watorekfutnet2022}.

Apart from the correlations in time, the asset-asset cross-correlations play an important role in shaping of the financial market structure as they lead to the emergence of the hierarchical organization of the markets as well as coupling between different markets~\cite{DrozdzS-2001a,Plerou2002,MaslovS-2001a,Crane2022}. While the hierarchical cross-correlations among the assets traded on the same market are a clear indicator of market maturity, the role of potential couplings between different markets must be interpreted with care. This is because either an independent dynamics of a market or a profound coupling of a market with the world's leading markets, being the two opposite cases, can potentially be interpreted in favour of market maturity. The former one because the independence can be viewed as strength and as a possibility to use the assets traded on such a market as a safe haven in hedging strategies~\cite{James2022inf,James2023ec}, and the latter one because it suggests that such a market is a well-rooted part of the global financial markets. However, intuitively, neither of these extremes seems to represent the notion of maturity well enough. It is more justified to view market maturity as an ability to switch its dynamics between independence and compliance because such a behaviour can better reflect the complexity that one may expect to be the property characterizing a developed market. This is why neither the effect of the cryptocurrency market decoupling from Forex reported in~\cite{DrozdzS-2019a} nor the effects of the cryptocurrency market independence~\cite{Corbet2018,Wang2019,Shahzad2019,shahzad2019a,bouri2020} and strong coupling between the cryptocurrencies and the traditional financial markets reported in~\cite{DrozdzS-2020a,James2021,James2021b,WatorekM-2023a}, respectively, can alone be a signature of maturity. It is rather the opposite: only such a flexible dynamics swinging between idiosyncrasy and strong subjugation of the market to an actual global trend can be a manifestation of market maturity.

In this work the stress is put on the investigation of the current statistical properties of the cryptocurrency log-returns and volume from the perspective of how these properties differ from their counterparts in the traditional financial markets: the stock markets, Forex, and the commodity markets. One has to be aware, though, that the statistical approach constitutes only a segment of the issues related to market maturity.

\section{Methods and results}

\subsection{Empirical dataset}

The data set studied contains 1 min quotations of 70 cryptocurrencies that were among the most actively traded on the Binance exchange~\cite{Binance}, which had the largest market share in 2022~\cite{marketshare}, over the period from Jan 1, 2020 to Dec 31, 2022 (3 years). The quotes are expressed in USD Tether (USDT), a stablecoin linked to the US dollar and its value is close to 1 USD by design~\cite{tether}. Basic time series statistics corresponding to these 70 cryptocurrencies are collected in Table~\ref{tab::data.stats}. For a time series of price quotations $Q(t_i)$, $i=1,...,T$, the equally spaced logarithmic returns $R_{\Delta t}(t_i) = \log Q(t_{i}) - \log Q(t_{i-1})$, where $t_{i}-t_{i-1}=\Delta t$, are derived. Fig.~\ref{fig::returns.integrated} shows the evolution of the cumulative log-returns $\hat{R}_{\Delta t}(t_i)=\sum_{i=1}^i R_{\Delta t}(t_i)$ during the whole period covered by the data. In accordance with the actual cryptocurrency price quotes, in 2021 the whole market experienced a transition from the bull phase to the bear phase.

\begin{table}[H]
\begin{adjustwidth}{-\extralength}{0cm}
\caption{Basic statistics of the cryptocurrencies considered in this study: the average inter-transaction time $\delta t$, the fraction of zero returns in time series \%0, the average volume value traded per minute $W$, and market capitalization $C$ on Jan 1, 2023. For the cryptocurrency name list see Table~\ref{tab::ticker.list} in Appendix~\ref{sect::appendix}.}
\label{tab::data.stats}
\centering
\begin{tabular}{|l|c|c|c|c||l|c|c|c|c|}
\hline
Ticker & $\delta t$ [$s$] & $\%0$ & $W$ [USDT] & $C$ [$\times 10^6$ USD] & Ticker & $\delta t$ [$s$] & $\%0$ & $W$ [USDT] & $C$ [$\times 10^6$ USD] \\ \hline\hline
BTC         & 0.04                   & 0.003           & 1,683,710               & 320,025                 & LINK        & 0.41                   & 0.095           & 84,423                 & 2,856                   \\ \hline
ADA         & 0.24                   & 0.121           & 172,891                & 8,621                   & LTC         & 0.41                   & 0.142           & 80,441                 & 5,096                   \\ \hline
ALGO        & 0.78                   & 0.117           & 24,320                 & 1,267                   & MATIC       & 0.32                   & 0.166           & 100,100                & 6,638                   \\ \hline
ANKR        & 1.84                   & 0.195           & 10,762                 & 151                    & MFT         & 5.01                   & 0.425           & 2,436                  & 54                     \\ \hline
ARPA        & 2.75                   & 0.165           & 6,082                  & 33                     & MTL         & 3.16                   & 0.400           & 5,122                  & 46                     \\ \hline
ATOM        & 0.58                   & 0.109           & 42,048                 & 2,710                   & NEO         & 1.45                   & 0.194           & 18,893                 & 451                    \\ \hline
BAND        & 2.13                   & 0.175           & 8,285                  & 49                     & NKN         & 2.99                   & 0.425           & 5,807                  & 56                     \\ \hline
BAT         & 1.53                   & 0.162           & 10,543                 & 251                    & NULS        & 4.44                   & 0.442           & 2,845                  & 12                     \\ \hline
BCH         & 0.70                   & 0.140           & 48,288                 & 1,869                   & OMG         & 0.83                   & 0.178           & 24,235                 & 146                    \\ \hline
BEAM        & 5.30                   & 0.433           & 2,089                  & 14                     & ONE         & 0.97                   & 0.227           & 21,983                 & 133                    \\ \hline
BNB         & 0.17                   & 0.095           & 276,261                & 39,052                  & ONG         & 5.53                   & 0.482           & 2,297                  & 71                     \\ \hline
CELR        & 1.77                   & 0.292           & 10,843                 & 68                     & ONT         & 1.28                   & 0.149           & 16,136                 & 134                    \\ \hline
CHZ         & 0.59                   & 0.232           & 51,827                 & 672                    & PERL        & 5.00                   & 0.431           & 2,406                  & 7                      \\ \hline
COS         & 2.63                   & 0.455           & 3,575                  & 18                     & QTUM        & 1.58                   & 0.179           & 14,178                 & 196                    \\ \hline
CTXC        & 3.42                   & 0.464           & 3,942                  & 33                     & REN         & 2.72                   & 0.207           & 6,232                  & 62                     \\ \hline
DASH        & 1.44                   & 0.206           & 14,543                 & 468                    & RLC         & 2.80                   & 0.293           & 6,090                  & 95                     \\ \hline
DENT        & 1.24                   & 0.353           & 16,417                 & 68                     & RVN         & 1.82                   & 0.202           & 9,699                  & 232                    \\ \hline
DOCK        & 5.39                   & 0.455           & 2,135                  & 12                     & STX         & 4.42                   & 0.416           & 3,847                  & 288                    \\ \hline
DOGE        & 0.20                   & 0.173           & 247,343                & 9,317                   & TFUEL       & 2.09                   & 0.353           & 10,411                 & 189                    \\ \hline
DUSK        & 2.97                   & 0.441           & 3,994                  & 34                     & THETA       & 0.64                   & 0.173           & 35,023                 & 733                    \\ \hline
ENJ         & 1.17                   & 0.225           & 21,114                 & 243                    & TOMO        & 3.84                   & 0.316           & 3,581                  & 24                     \\ \hline
EOS         & 0.53                   & 0.147           & 59,616                 & 948                    & TROY        & 3.20                   & 0.381           & 3,347                  & 23                     \\ \hline
ETC         & 0.58                   & 0.099           & 63,736                 & 2,188                   & TRX         & 0.46                   & 0.142           & 71,306                 & 5,041                   \\ \hline
ETH         & 0.10                   & 0.010           & 853,284                & 146,967                 & VET         & 0.52                   & 0.093           & 55,362                 & 1,163                   \\ \hline
FET         & 2.65                   & 0.255           & 7,909                  & 75                     & VITE        & 4.22                   & 0.469           & 3,078                  & 18                     \\ \hline
FTM         & 0.50                   & 0.174           & 63,723                 & 556                    & WAN         & 7.24                   & 0.303           & 1,609                  & 34                     \\ \hline
FUN         & 3.91                   & 0.538           & 2,911                  & 66                     & WAVES       & 1.19                   & 0.177           & 19,265                 & 144                    \\ \hline
HBAR        & 1.57                   & 0.268           & 11,765                 & 957                    & WIN         & 1.01                   & 0.283           & 26,244                 & 72                     \\ \hline
HOT         & 0.96                   & 0.237           & 22,543                 & 250                    & XLM         & 0.78                   & 0.165           & 33,309                 & 1,894                   \\ \hline
ICX         & 2.64                   & 0.306           & 6,951                  & 135                    & XMR         & 1.62                   & 0.184           & 14,164                 & 2,707                   \\ \hline
IOST        & 1.40                   & 0.199           & 14,551                 & 129                    & XRP         & 0.21                   & 0.071           & 229,976                & 17,055                  \\ \hline
IOTA        & 1.53                   & 0.168           & 12,077                 & 478                    & XTZ         & 1.08                   & 0.137           & 19,407                 & 663                    \\ \hline
IOTX        & 1.52                   & 0.266           & 11,894                 & 203                    & ZEC         & 1.15                   & 0.240           & 20,010                 & 597                    \\ \hline
KAVA        & 1.57                   & 0.155           & 12,888                 & 198                    & ZIL         & 1.03                   & 0.145           & 20,195                 & 258                    \\ \hline
KEY         & 2.83                   & 0.358           & 4,310                  & 15                     & ZRX         & 3.04                   & 0.214           & 5,674                  & 128                    \\ \hline
\end{tabular}
\end{adjustwidth}
\end{table}

\subsection{Cumulative distribution functions of returns and volume}

\begin{figure}[ht!]

\includegraphics[width=1\textwidth]{Figs/returnsintegrated.eps}
\caption{Evolution of the cumulative log-returns $\hat{R}(t)$ of the 70 cryptocurrencies over the time period from 1 January 2020 to 31 December 2022. The~colors of two of the most liquid cryptocurrencies and a few other distinguished ones are indicated explicitly. The~bulk of the cryptocurrencies is shown in the background (grey lines).}
\label{fig::returns.integrated}
\end{figure}

The cumulative distribution function (cdf) $P(X>r_{\Delta t})$ can be calculated from the normalized returns $r_{\Delta t}(t_i)=(R_{\Delta t}(i)-\mu)/ \sigma$, with $\mu$ and $\sigma$ denoting sample mean and standard deviation, respectively. A form of this distribution varies among the markets and assets, but some interesting properties can be observed. There are generally three factors that shape it: the first one is liquidity, the second one is trading speed, and the third one is the overall market volatility~\cite{Farmer2004}. If one focuses on a specific market, the most liquid assets show faster decline of $P(X>r_{\Delta t})$ with $r_{\Delta t}$ than the less liquid ones for a given $\Delta t$~\cite{Drozdz2007}. However, most of the assets traded on mature markets reveal a power-law dependence of $P(X>r_{\Delta t})$ for some range of $\Delta t$~\cite{Plerou1999a,Gopikrishnan1999,Gabaix2003,Drozdz2003,Drozdz2007}:
\begin{equation}
P(X>r_{\Delta t}) \sim |r_{\Delta t}|^{-\gamma},
\label{eq::inverse.cubic}
\end{equation} 
with $\gamma \approx 3$. It is observed for short sampling intervals and it is persistent for a range of $\Delta t$ due to the existence of strong inter-asset correlations. This inverse cubic power-law dependence breaks for sufficiently long $\Delta t$ and the cdf tails converge to the expected normal distribution. The speed of information processing on a given market also has its influence on the crossover $\Delta t$. Since this speed increases with time as new technologies enter the service, we observe a gradual decrease of the crossover $\Delta t$ across decades. The speed of market trading allows for a larger transaction number in time unit, so this factor accelerates the market time even more~\cite{Drozdz2007}. The emerging markets where investment strategies require the accommodation of significant risk are thus highly volatile. The cdfs of the asset returns in this case often show heavy tails with the scaling exponent $\gamma \ll 3$, sometimes even in the L\'evy-stable regime. In such markets the inverse cubic behaviour of $P(X>r_{\Delta t})$ may occur for some assets only, while for the other assets it cannot be found at all. This is why such extreme tails are often considered to be an indicator of market immaturity~\cite{DrozdzBTC2018}. 

\begin{figure}[ht!]
\begin{adjustwidth}{-\extralength}{0cm}

\includegraphics[width=0.67\textwidth]{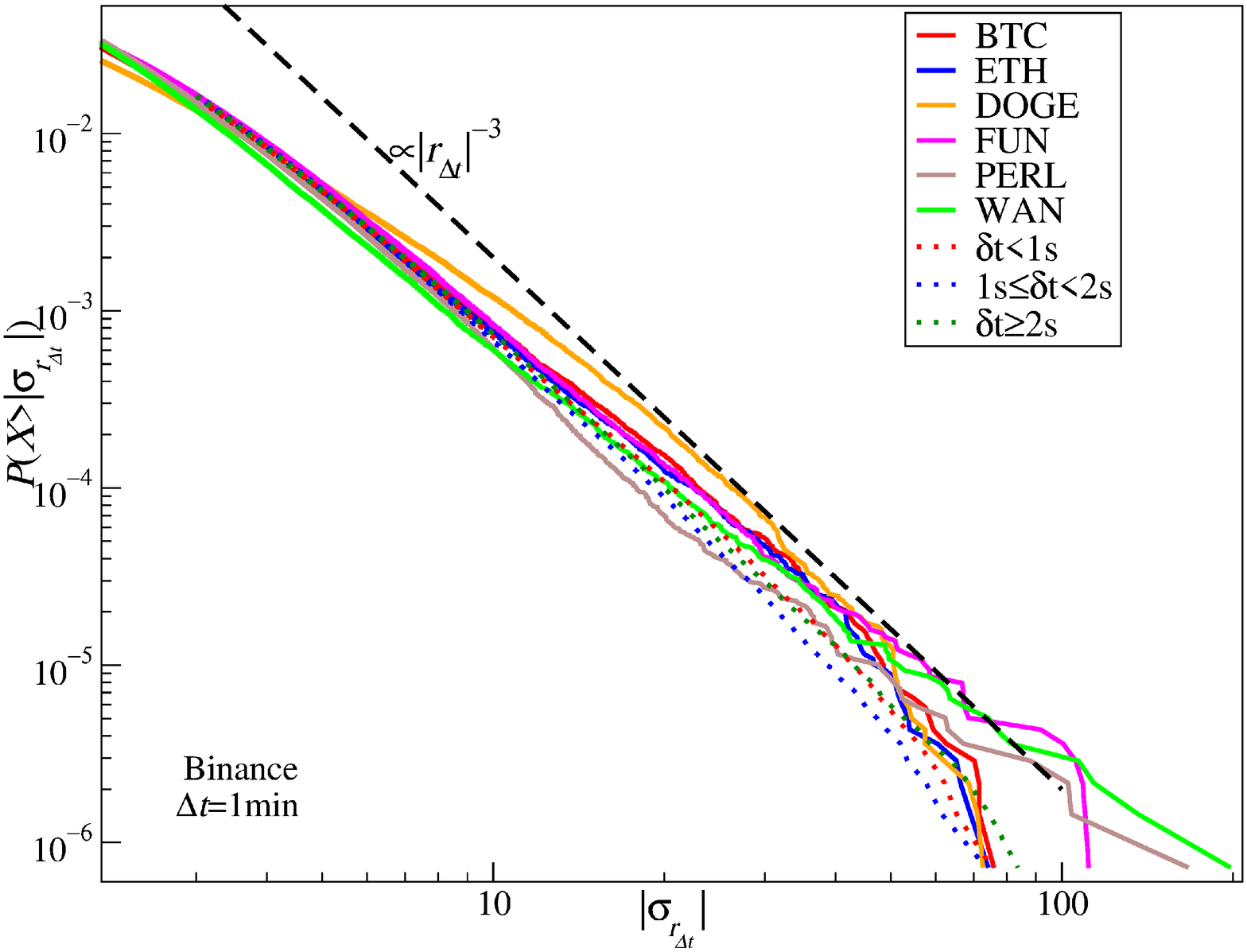}
\hspace{0.5cm}
\includegraphics[width=0.67\textwidth]{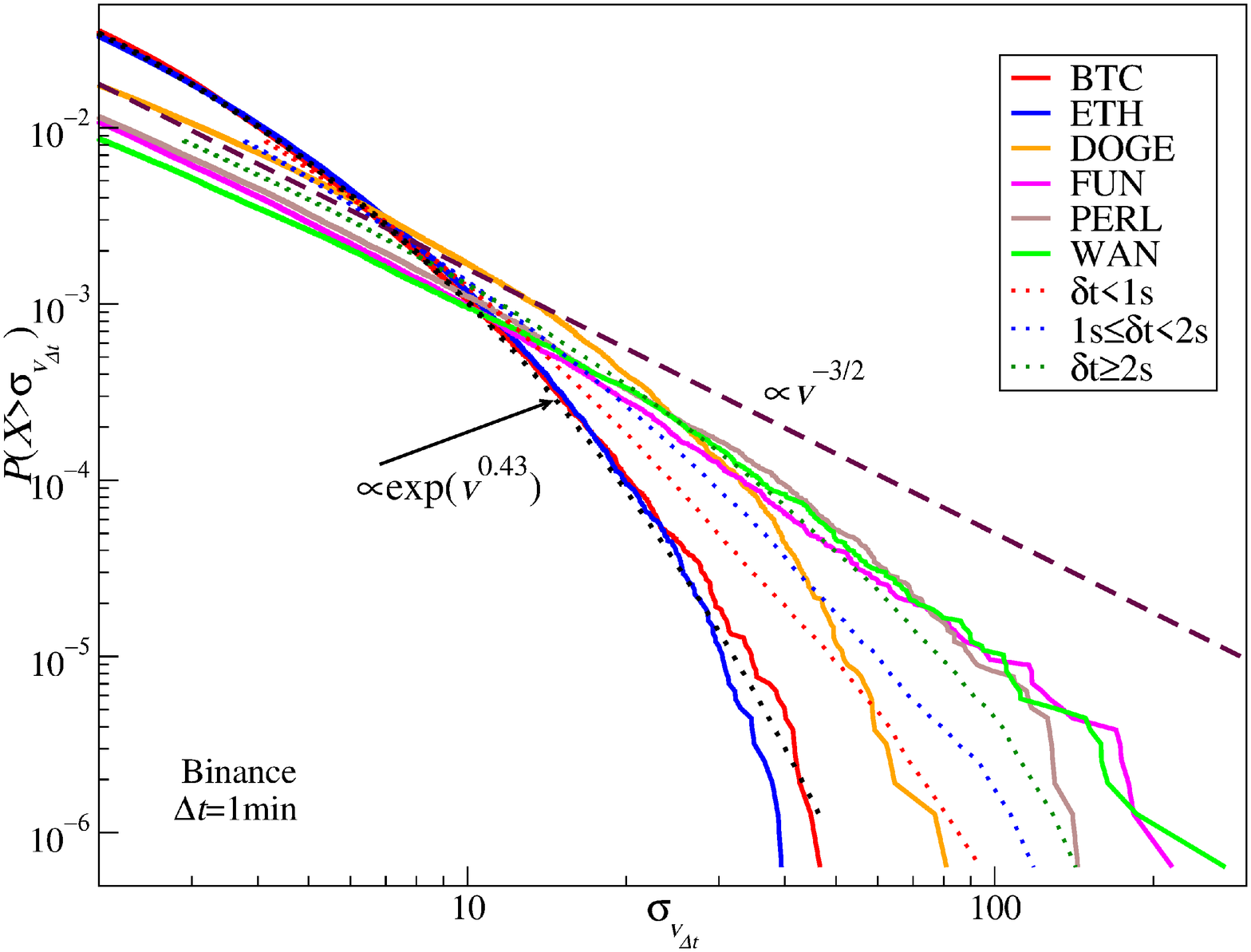}
\end{adjustwidth}
\caption{Cumulative distribution functions of the absolute normalized log-returns $r_{\Delta t}$ (left) and the normalized volume traded $v_{\Delta t}$ (right) for $\Delta t=1$ min in units of the respective standard deviations $\sigma$ for the selected cryptocurrencies with the highest liquidity (BTC and ETH) or the heaviest tails (DOGE, FUN, PERL, and WAN). The~average cumulative distribution functions for the cryptocurrencies with the average inter-transaction time fulfilling the relations $\delta t<1s$ (Group I, dotted red), $1s\le \delta t<2s$ (Group II, dotted blue), and~$\delta t \ge 2s$ (Group III, dotted green) are also shown. Power laws with the scaling exponents $\gamma$ and $\beta$ assuming values typical for the financial markets---$\gamma = 3$ and $\beta=3/2$---are denoted by dashed lines. There is also a stretched exponential function fitted to the $v_{\Delta t}$ distributions for BTC and ETH on the right (black dotted line).}
\label{fig::return.volume.cdf}

\end{figure}

Based on the average inter-transaction time $\delta t$, we categorize the considered cryptocurrencies into 3 Groups: I - the most frequently traded ones ($\delta t<1s$), II - the ones with the average trading frequency ($1s\le \delta t<2s$), and III - the least frequently traded ones ($\delta t \ge 2s$). Then we calculate the average cdfs for the cryptocurrencies belonging to each group. We show these cdfs in Fig.~\ref{fig::return.volume.cdf} (left panel, dotted lines) together with the cdfs for a few selected individual cryptocurrencies (solid lines). Their form can be compared with the inverse cubic power-law model denoted by a dashed line. It occurs that the average distributions have their tail that is close to a power law with the exponent $\gamma$ being close to 3. The most liquid cryptocurrencies - BTC and ETH - develop tails that show a cross-over from the power-law regime to CLT-like regime for the relatively small values of $|r_{\Delta t}|$ as compared to both the average cdfs and to less frequently traded individual cryptocurrencies like FUN, PERL, and WAN. The case of Dogecoin, which has the smallest slope in the middle of the distribution and, at the same time, his tail is not the thickest, is special. On the one hand, it can be included among the main cryptocurrencies, due to the high frequency of transactions and capitalization, and at the same time, it was the subject of possible price manipulation through Elon Musk's tweets~\cite{Nani2022,SHAHZAD2022Doge}.

Another quantity that is frequently observed to be power-law distributed is normalized volume traded in time unit $v_{\Delta t}(t_i)=(V_{\Delta t}(i)-\mu)/ \sigma$~\cite{GopikrishnanP-2000a,Gabaix2003}: 

\begin{equation}
P(X>v_{\Delta t}) \sim v_{\Delta t}^{-\beta}.
\label{eq::volume.traded}
\end{equation} 
In this case, the exponent is much lower than for the absolute returns and corresponds to the L\'evy-stable regime: $\beta < 2$. It was argued that there exists a simple relation between both the exponents: $\beta = \gamma / 2$~\cite{Gabaix2003}. Fig.~\ref{fig::return.volume.cdf} (right panel) shows the cumulative distribution functions for $v_{\Delta t}$ for the same individual cryptocurrencies and their Groups I-III as in Fig.~\ref{fig::return.volume.cdf} (left panel). Now the cdfs for BTC and ETH do not develop power-law tails. A model that fits them the best is the stretched exponential function: $P(X>v_{\Delta t}) \sim \exp \sigma_v^{-\eta}$ with $\eta = 0.43$. However, in the case of less frequently traded cryptocurrencies, which belong to Group III, one can observe the power-law relation. What makes the results obtained here different from their counterparts for, for instance, the stock markets is that one does not find any cryptocurrency with its cdf being power law with the exponent 3/2; the cdf tails decrease considerably faster here.

\subsection{Price impact}

At this point, it is worthwhile to consider a possible causal relation between the returns and the volume despite the fact that no clear relation can be seen between their cdfs. It revokes the empirically well-documented observation that volume can influence price changes (both on the level of the order book and the level of the aggregated transaction volume), which is known in the literature as the price impact~\cite{BouchaudJP-2010a,DufourA-2000a,Gabaix2003,WeberP-2005a,WilinskiM-2014a,ContR-2014a}. In order to investigate this issue, for each cryptocurrency, two parallel time series corresponding to $|R_{\Delta t}(t)|$ and $V_{\Delta t}(t)$ are input to the $q$-dependent detrended cross-correlation coefficient $\rho_q$ measuring how correlated are two detrended residual signals across different scales~\cite{kwapien2015}. Definition of the coefficient $\rho_q$, which allows one to quantify cross-correlations between two nonstationary signals, is based on the multifractal detrended cross-correlation analysis (MFCCA), whose algorithm can be sketched as follows~\cite{oswiecimka2014}. 

In this particular case, there are two time series of length $T$ and sampling intervals $\Delta t$: $|R_{\Delta t}(t_i)|$ and $V_{\Delta t}(t_i)$ with $i=1,...,T$. One starts the procedure by dividing each time series into $M_s=2 \lfloor T/s \rfloor$ non-overlapping segments of length $s$ (called \textit{scale}) going from both ends ($\lfloor \cdot \rfloor$ denotes floor value). In each segment labelled by $\nu$, both signals are integrated and polynomial trends $P_{\cdot,s,\nu}^{(m)}$ of degree $m$ are removed:
\begin{align}
\hat{R}_{\Delta t}(t_j,s,\nu) & = \sum_{k=1}^j |R_{\Delta t}(t_{s(\nu-1)+k})| - P_{R,s,\nu}^{(m)}(t_j),\\
\hat{V}_{\Delta t}(t_j,s,\nu) & = \sum_{k=1}^j V_{\Delta t}(t_{s(\nu-1)+k}) - P_{V,s,\nu}^{(m)}(t_j),
\label{eq::detrended.profile}
\end{align}
where  $j=1,...,s$ and $\nu=1,...,M_s$. The detrended covariance is derived as
\begin{equation}
f_{|R|V}^2(s,\nu) = {1 \over s} \sum_{j=1}^s \left[ \hat{R}_{\Delta t}(t_j,s,\nu) - \langle \hat{R}_{\Delta t}(t_j,s,\nu) \rangle_j \right] \left[ \hat{V}_{\Delta t}(t_j,s,\nu) - \langle \hat{V}_{\Delta t}(t_j,s,\nu) \rangle_j \right],
\label{eq::detrended.covariance}
\end{equation}
where $\langle \cdot \rangle_j$ denotes the averaging over $j$. The detrended covariances calculated for all the segments $\nu$ are then used to determine the bivariate fluctuation function~\cite{oswiecimka2014}:
\begin{equation}
F_q^{|R|V}(s) = \big\{ {1 \over M_s} \sum_{\nu=1}^{M_s} {\rm sgn} [f_{|R|V}^2(s,\nu)] | f_{|R|V}^2(s,\nu)|^{q/2} \big\}^{1/q}.
\label{eq::fluctuation.function.rv}
\end{equation}
Apart from the bivariate form given by the formula above, the univariate fluctuation functions $F_q^{|R||R|}(s)$ and $F_q^{VV}(s)$ can also be calculated but in this case the covariance functions become variances and do not need to be factorized into the sign and modulus parts as no negative value can occur.

\begin{figure}[ht!]
\includegraphics[width=1\textwidth]{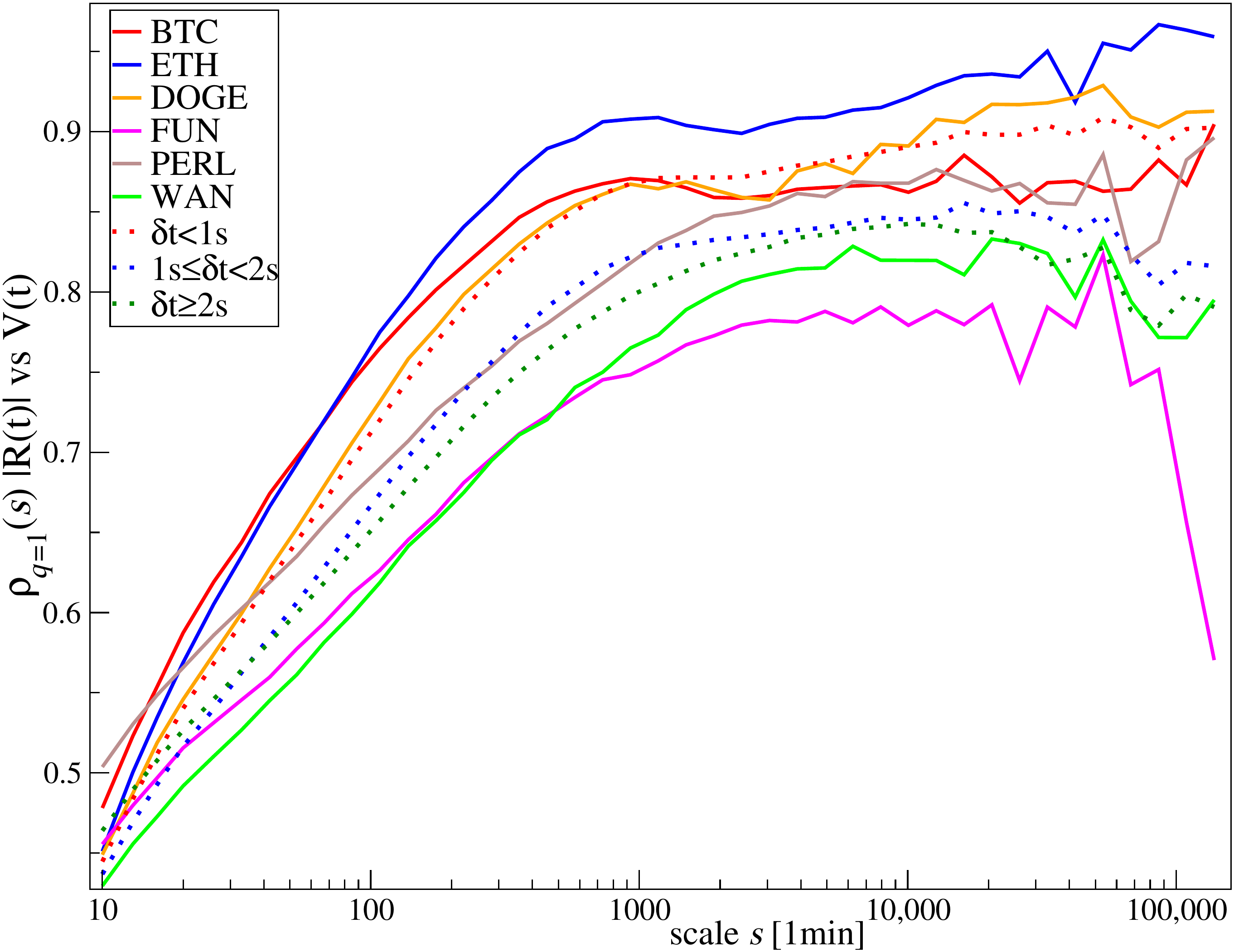}
\caption{The $q$-dependent detrended cross-correlation coefficient $\rho_q(s)$ of order $q=1$ calculated for volatility $|R_{\Delta t}(t)|$ and volume $V_{\Delta t}(t)$ (with $\Delta t=1$ min) for the selected individual cryptocurrencies---BTC, ETH, DOGE, FUN, PERL, and~WAN---where the cryptocurrency Groups I-III are characterized by a specific range of the average inter-transaction time: $\delta t<1s$ (Group I, dotted red), $1s \le \delta t<2s$ (Group II, dotted blue), $\delta t \ge 2s$ (Group III, dotted green). The~coefficient $\rho_q(s)$ has been averaged over all the cryptocurrencies belonging to a given~group.}
\label{fig::rho.q.rv}

\end{figure}

The above elements of the formalism allow one to introduce the $q$-dependent detrended cross-correlation coefficient $\rho_q(s)$ defined as~\cite{kwapien2015}:
\begin{equation}
\rho^{|R|V}_q(s) = {F_q^{|R|V}(s) \over \sqrt{ F_q^{|R||R|}(s) F_q^{VV}(s) }}.
\label{eq::rho.q}
\end{equation}
By manipulating value of the parameter $q$, one can focus on the correlations between fluctuations of different size: the large ones $q > 2$ or the small ones $q < 1$. For $q=2$ all the fluctuations in time series are considered with the same weights. For positive $q$, values of $\rho_q$ are restricted to the interval [-1,1] with their interpretation being similar to the interpretation of the classic Pearson coefficient $C$: $\rho_q=1$ means a perfect correlation, $\rho_q=0$ means independence, and $\rho_q=-1$ means a perfect anticorrelation. For negative $q$, the interpretation of the coefficient is more delicate and requires some experience~\cite{kwapien2015}. Fig.~\ref{fig::rho.q.rv} presents the coefficient $\rho_q(s)$ calculated in a broad range of scales $s$ for the selected individual cryptocurrencies (BTC, ETH, DOGE, FUN, PERL, and WAN) and the average $\rho_q(s)$ for Groups I-III. While different data sets are characterized by different strength of the detrended cross-correlations with Group I cross-correlated the strongest and Group 3 the weakest, there is an explicit division of scales into the short-scale range ($s < 1000$ min) where the correlations monotonously increase with increasing $s$ and the long-scale range ($s>1000$ min) where one observes a kind of saturation-like behaviour. In the latter, the correlations are characterized by $0.75 \le \rho_q(s) \le 0.95$, which means that the cryptocurrency market does not differ from other financial markets and its volatility $|R_{\Delta t}|$ and volume traded are strongly correlated. The two distinguished scale ranges are related to the information processing speed of the market: it requires some amount of time for the investors to fully react to the incoming information and to build up the cross-correlations. One might view this result as a counterpart of the Epps effect for the detrended volatility-volume data~\cite{EppsTW-1979a,kwapien2004,Toth2009,DrozdzS-2010a,watorek2021}. The main difference between this market and the regular financial markets is the relatively long cross-over scale ($s \approx 1000$ min), which can be associated with its worse liquidity.

\begin{figure}[ht!]
\begin{adjustwidth}{-\extralength}{0cm}

\includegraphics[width=0.67\textwidth]{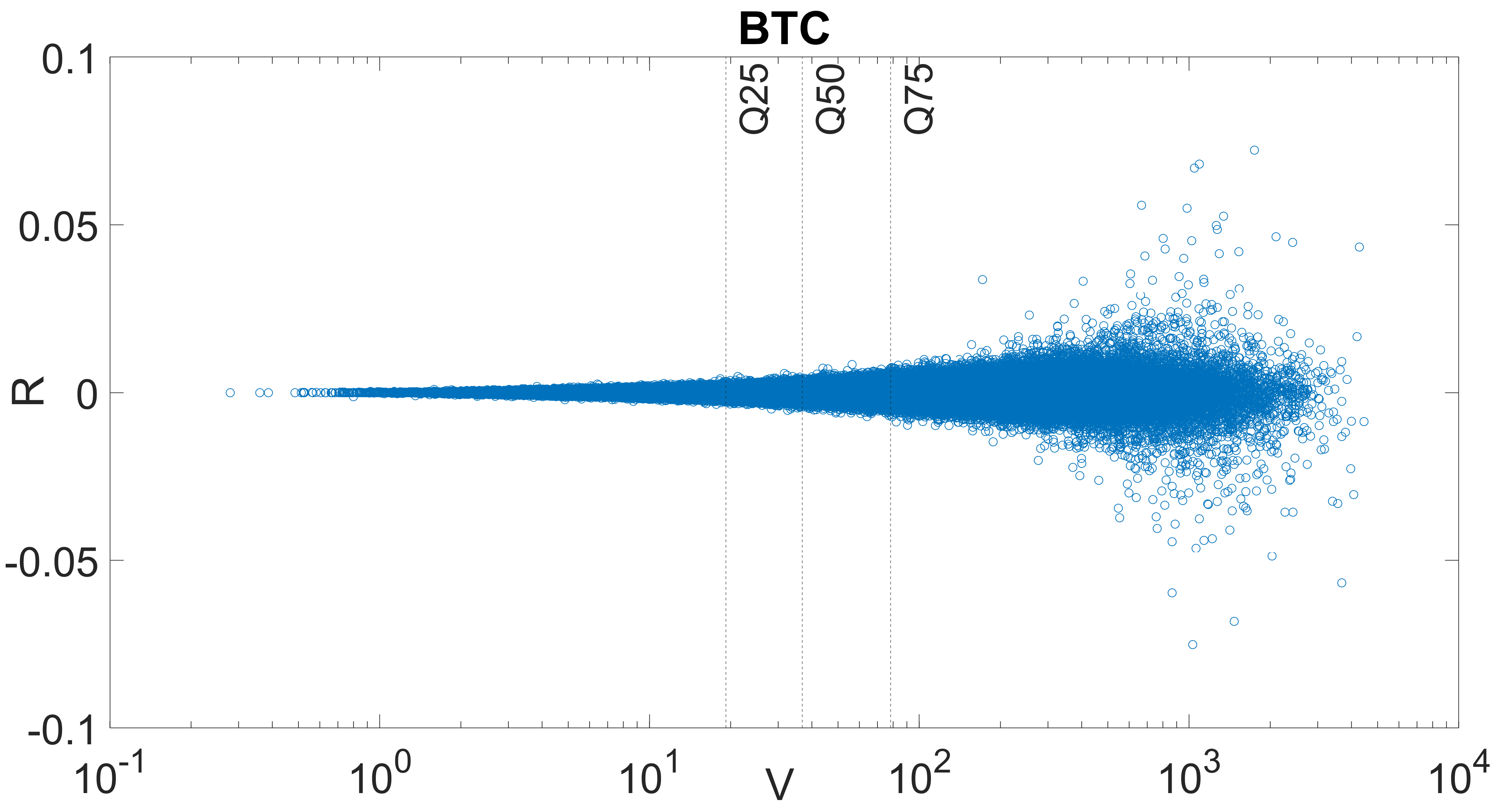}
\includegraphics[width=0.67\textwidth]{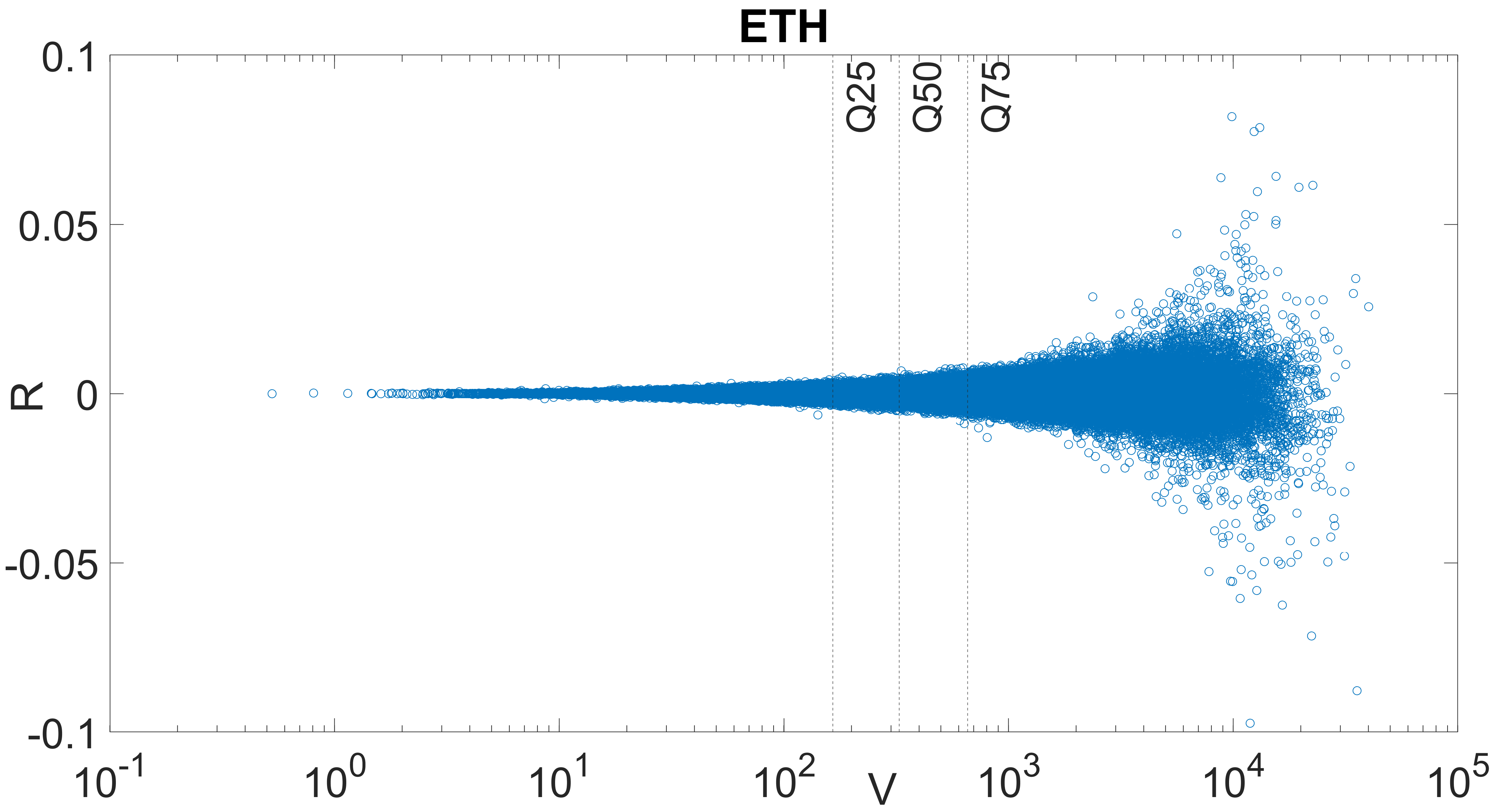}
\includegraphics[width=0.67\textwidth]{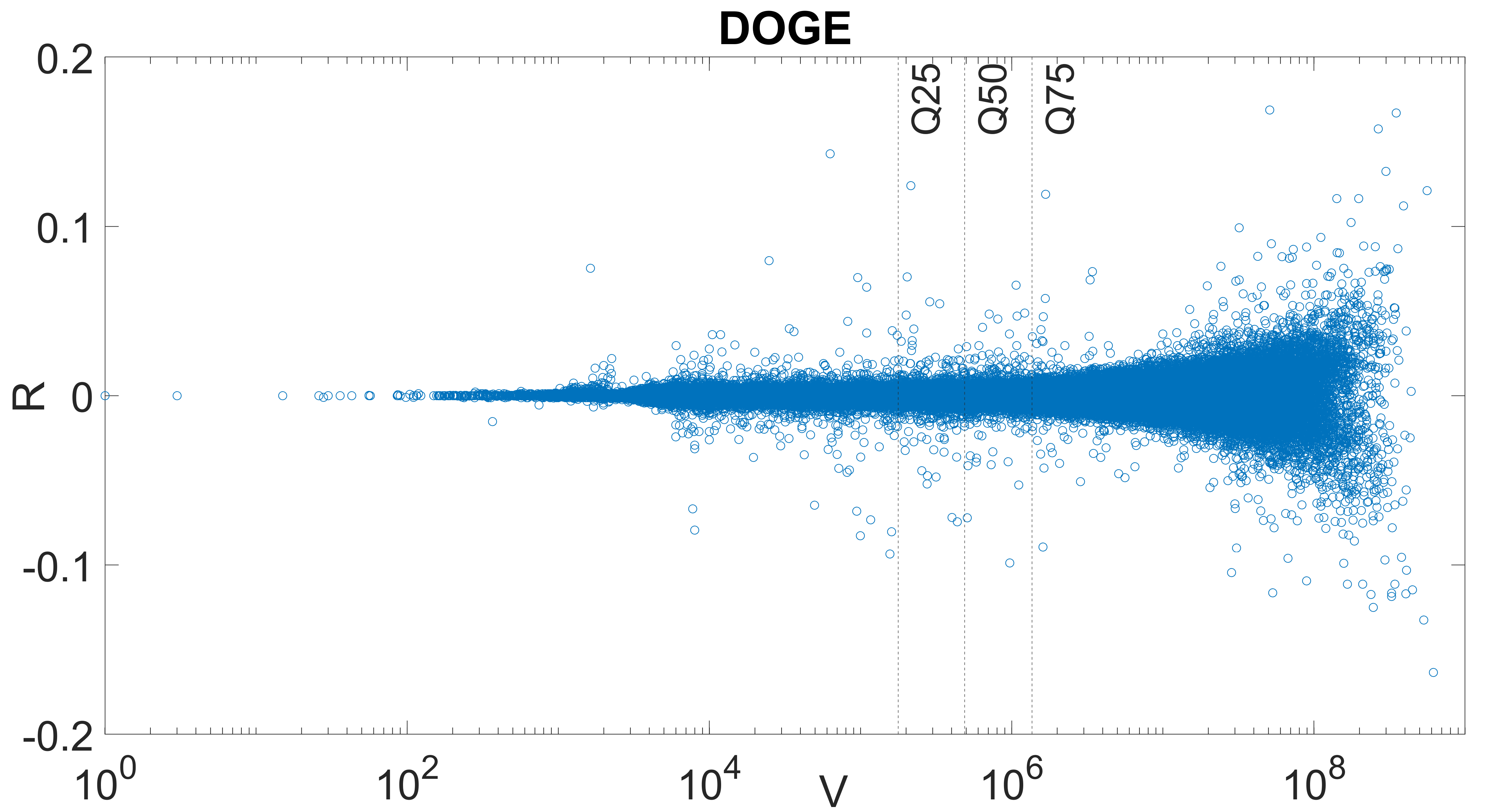}
\includegraphics[width=0.67\textwidth]{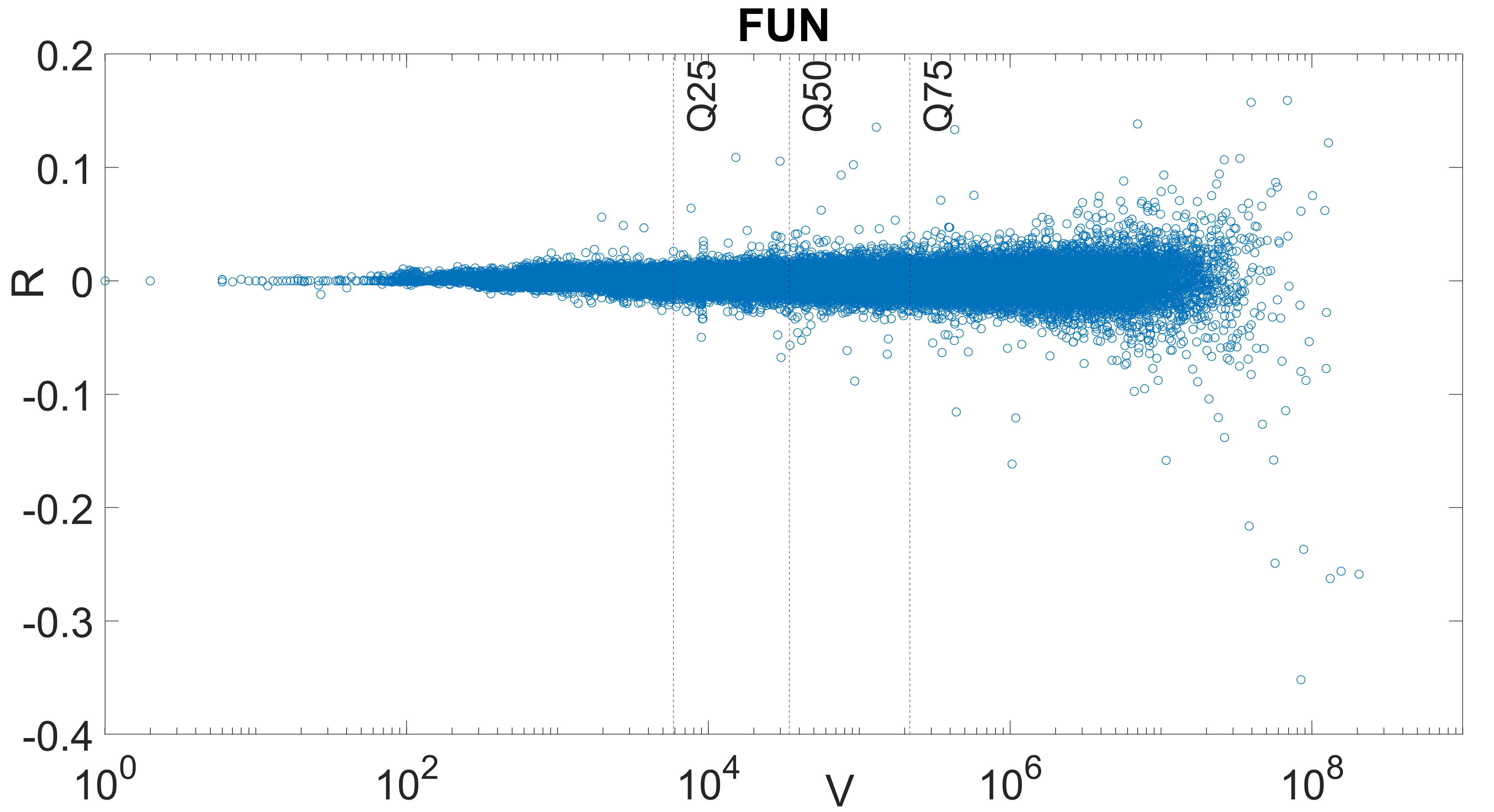}
\includegraphics[width=0.67\textwidth]{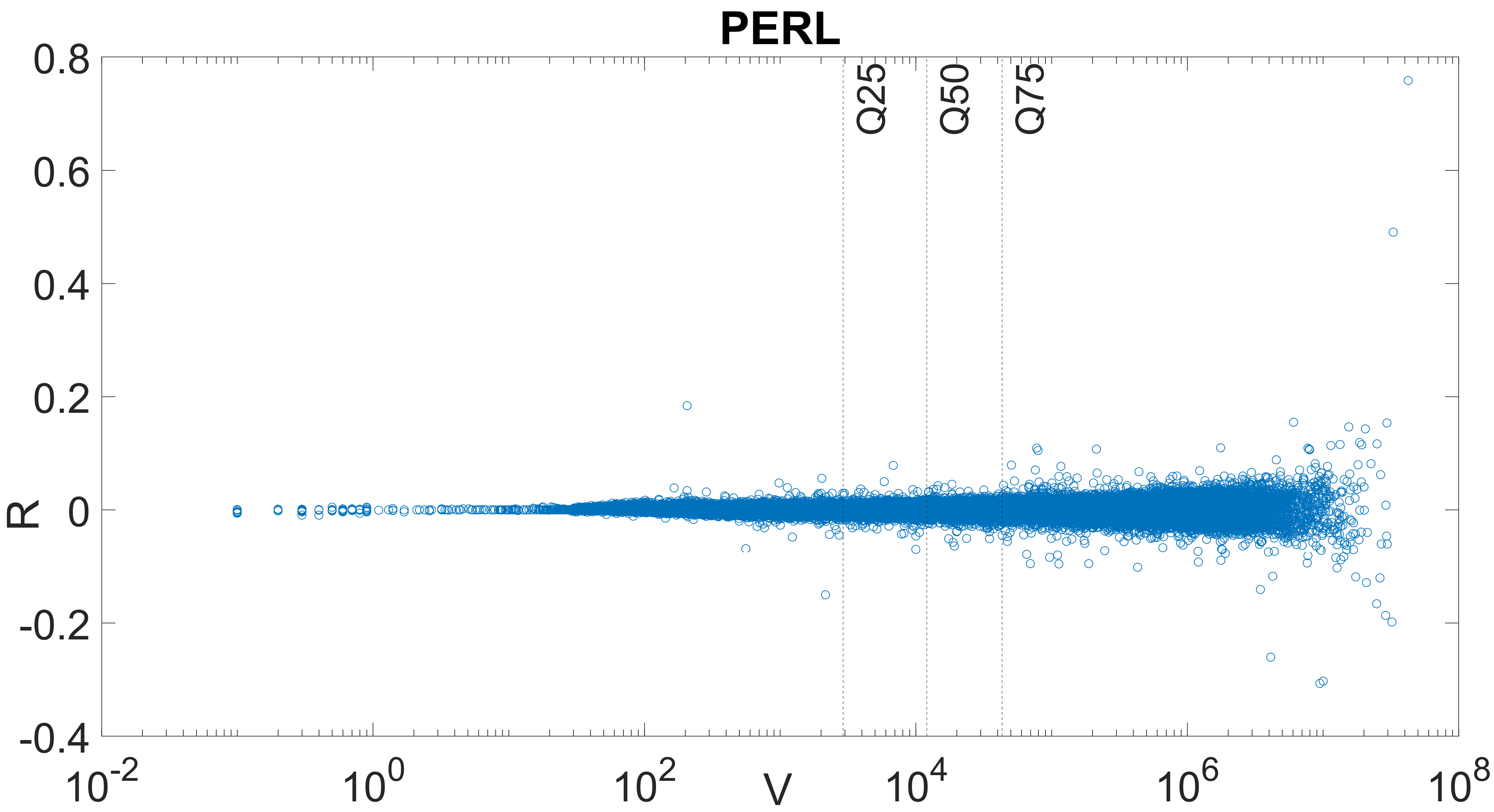}
\includegraphics[width=0.67\textwidth]{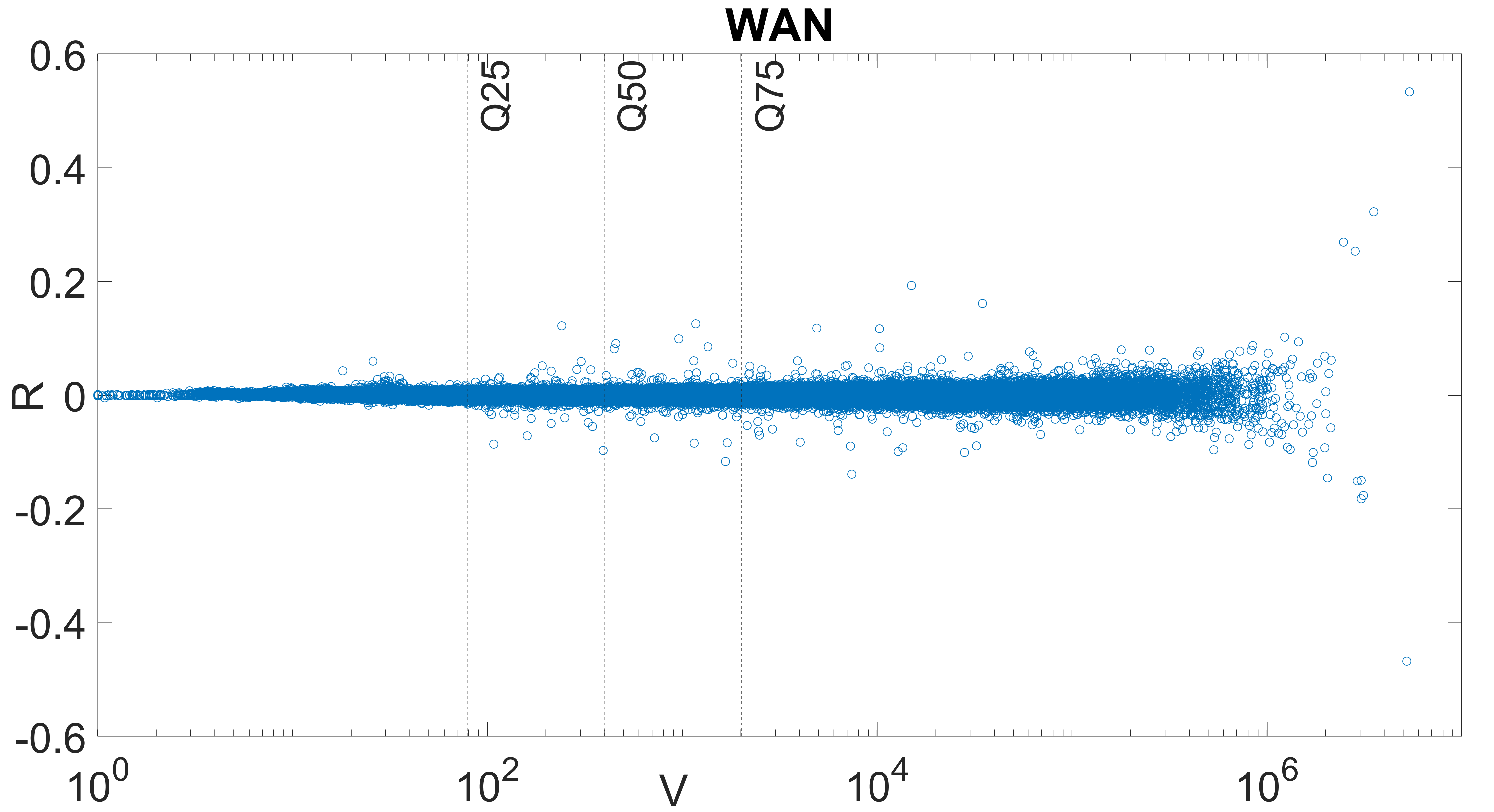}
\end{adjustwidth}
\caption{Scatter plots of the returns $R_{\Delta t}(t)$ and volume traded $V_{\Delta t}(t)$ for a few selected cryptocurrencies (BTC, ETH, DOGE, FUN, PERL, and~WAN). Each point corresponds to a specific 1 min long interval in the whole 3-year-long period of interest. The~vertical dashed lines in each panel denote the 25th, 50th, and~75th quantile of the volume probability distribution function for a given cryptocurrency. Note the logarithmic horizontal axis and the varying axis ranges among the~panels.}
\label{fig::scatterplots.rv}

\end{figure}

Next question to be asked is if there exists any functional relationship between $|R_{\Delta t}|$ and $V_{\Delta t}$. In order to address this question, the $R_{\Delta t}$ vs. $V{\Delta t}$ scatter plots for 6 selected cryptocurrencies have been created; see Fig.~\ref{fig::scatterplots.rv}. In general, the cross-correlations identified by means of $\rho_q(s)$ can also be confirmed visually on these plots: the larger the volume is, the larger can be the volatility. However, no specific functional form of $R_{\Delta t}(V_{\Delta t})$ can be inferred from this picture. Therefore, it is instructive to change the presentation to the conditional probability plots of the form $\mathbb{E}[f(|r_{\Delta t}|) | v_{\Delta t}]$, where the expectation value $\mathbb{E}[\cdot]$ can be approximated by the mean $\langle \cdot \rangle$. From a perspective of the market with the substantially limited liquidity, small price changes correspond to small transaction volumes and constitute market noise. Thus, one may expect that the most interesting relation between volatility and volume can be seen for large returns: $|r_{\Delta t}(t)| \gg 1$.

\begin{figure}[ht!]

\begin{adjustwidth}{-\extralength}{0cm}
\includegraphics[width=0.7\textwidth]{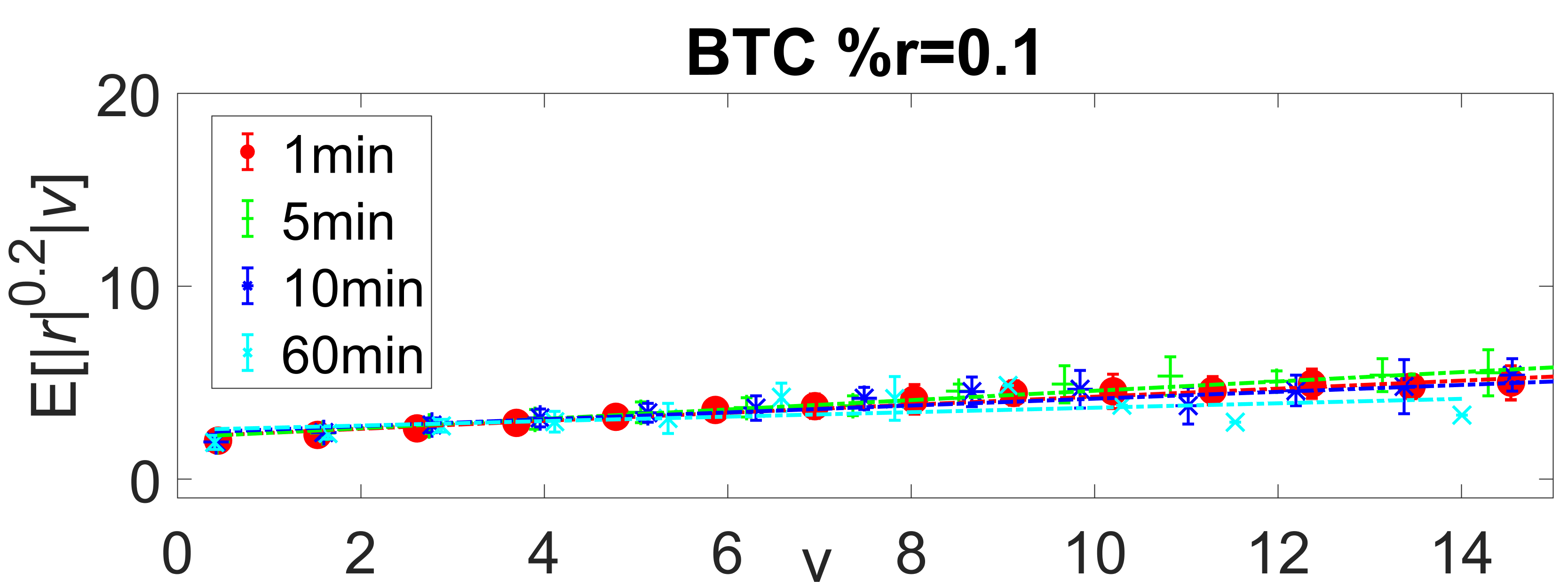}
\includegraphics[width=0.7\textwidth]{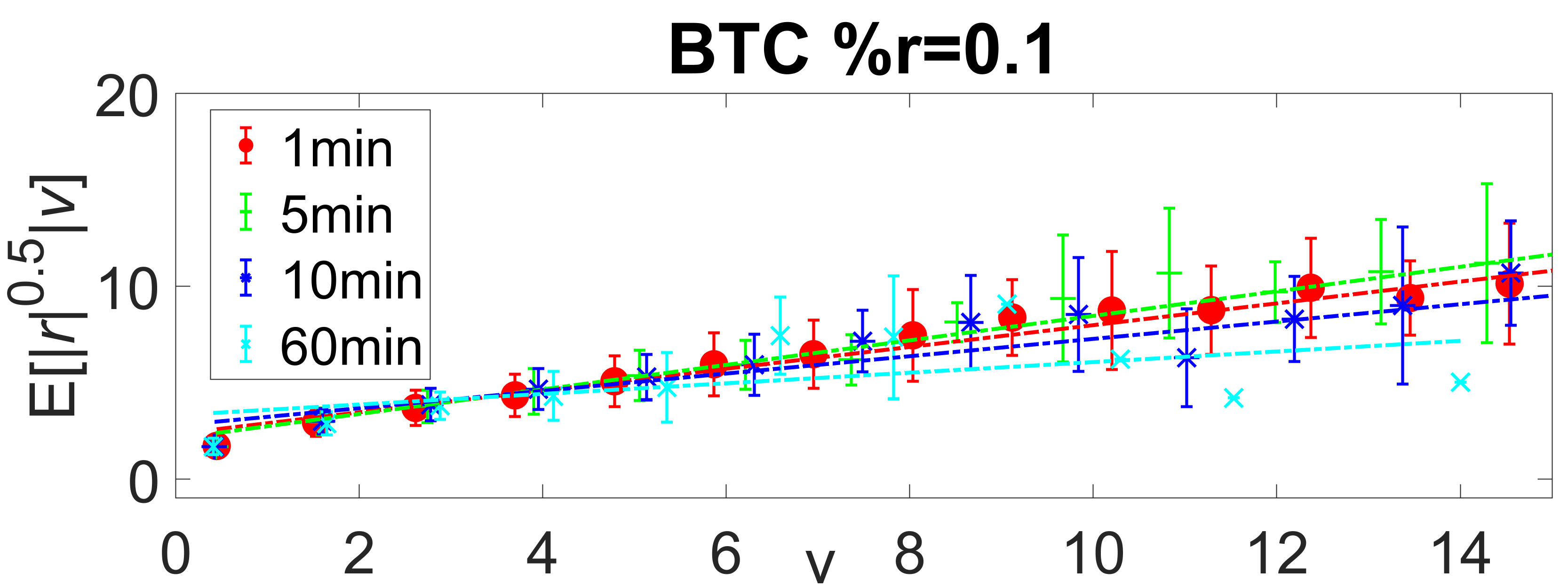}
\includegraphics[width=0.7\textwidth]{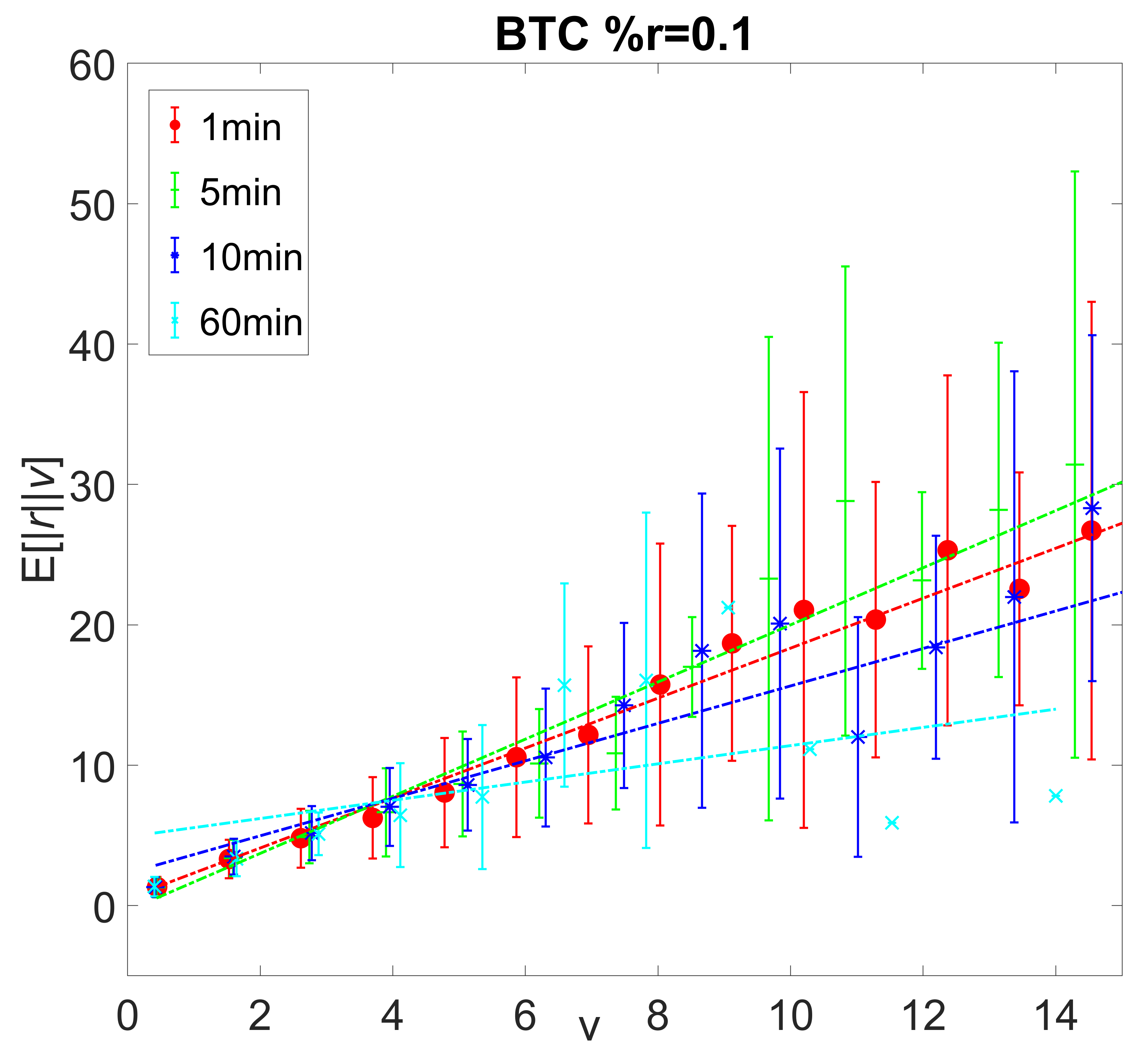}
\includegraphics[width=0.7\textwidth]{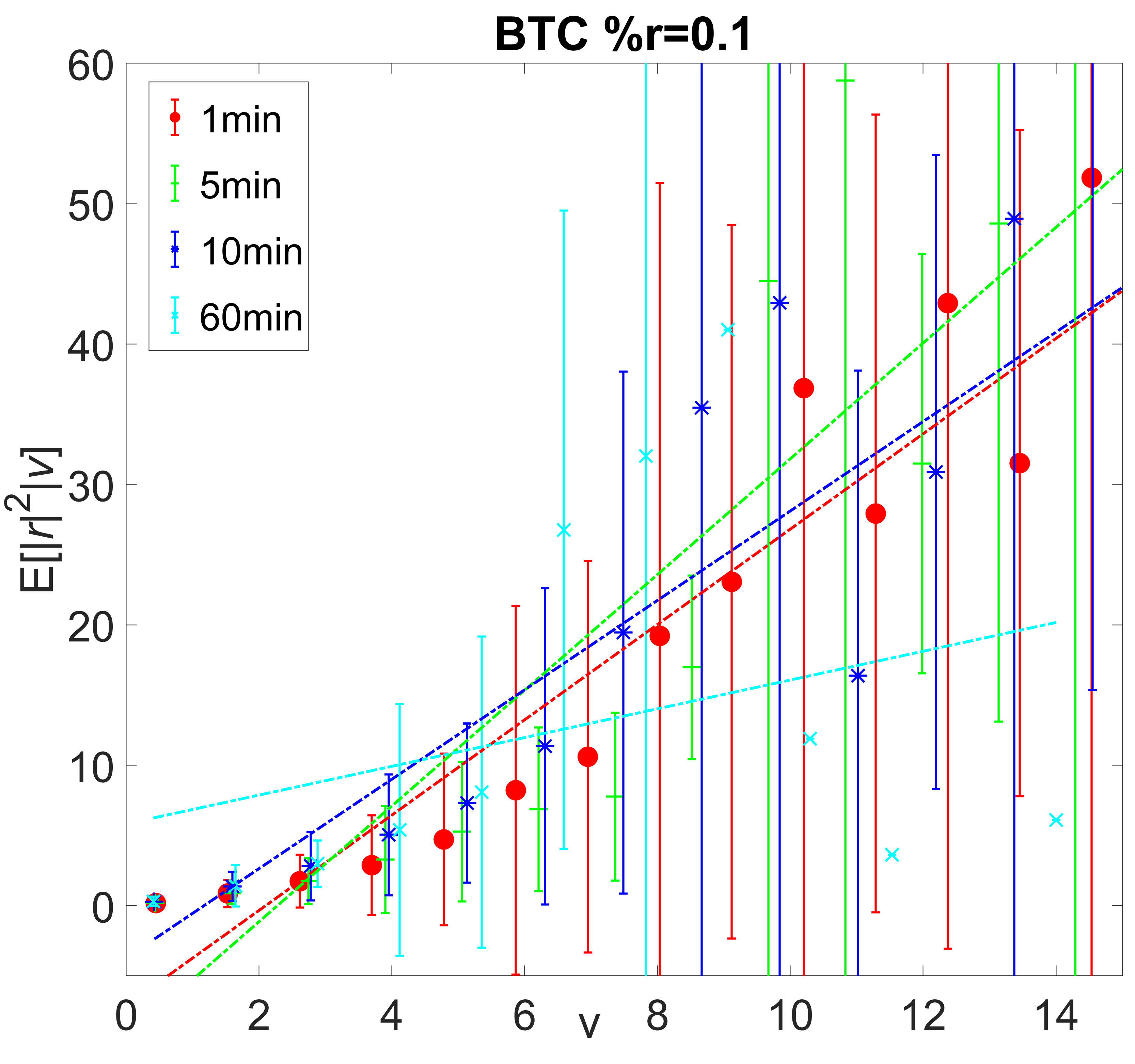}
\caption{Conditional expectation $\mathbb{E}[|r_{\Delta t}|^{\kappa}|v_{\Delta t}]$ for BTC if only a $p$-fraction of the largest normalized returns $r_{\Delta t}$ is preserved for each value of the normalized volume $v_{\Delta t}$. Each panel shows the results for a specific value of $\kappa$ together with a corresponding fitted power-law model. Four cases of the sampling interval are presented: $\Delta t=1$ min, 5 min, 10 min, and 60 min. The error bars show the conditional standard deviation $\sigma[|r_{\Delta t}|^{\kappa} | v_{\Delta t}]$.}
\label{fig::conditional.rv.BTC}
\end{adjustwidth}
\end{figure}

\begin{figure}[ht!]

\begin{adjustwidth}{-\extralength}{0cm}
\includegraphics[width=0.7\textwidth]{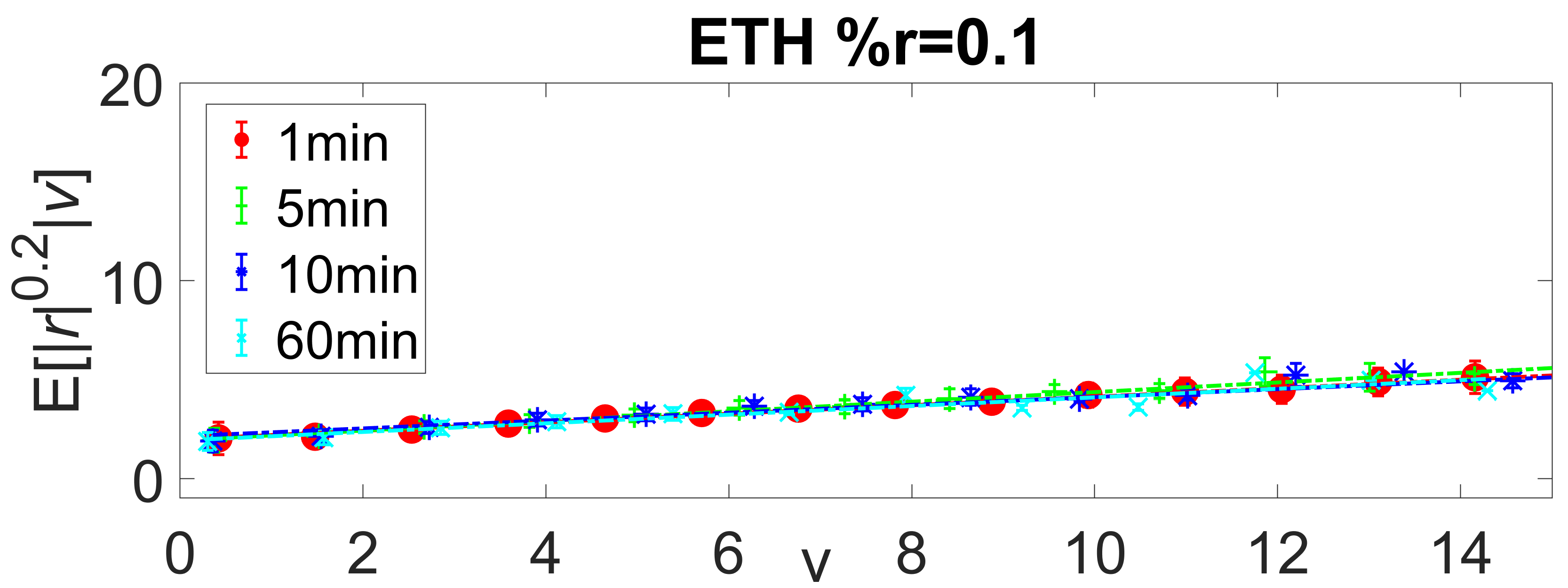}
\includegraphics[width=0.7\textwidth]{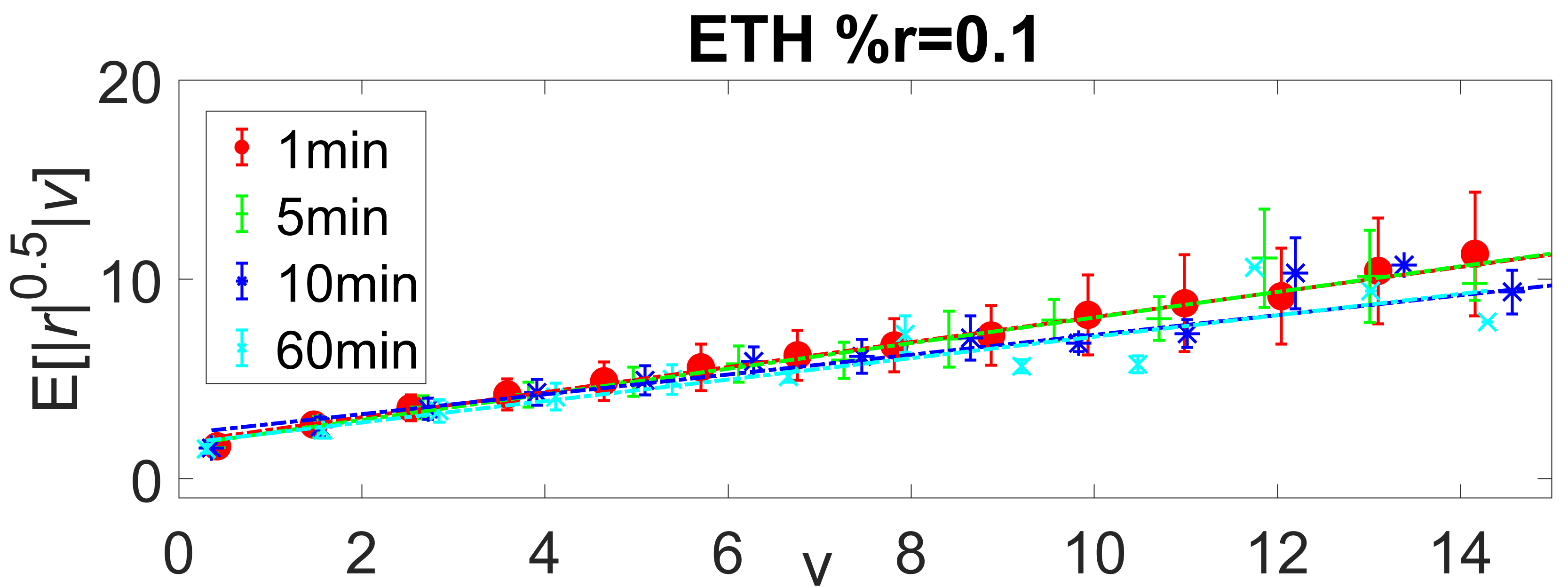}
\includegraphics[width=0.7\textwidth]{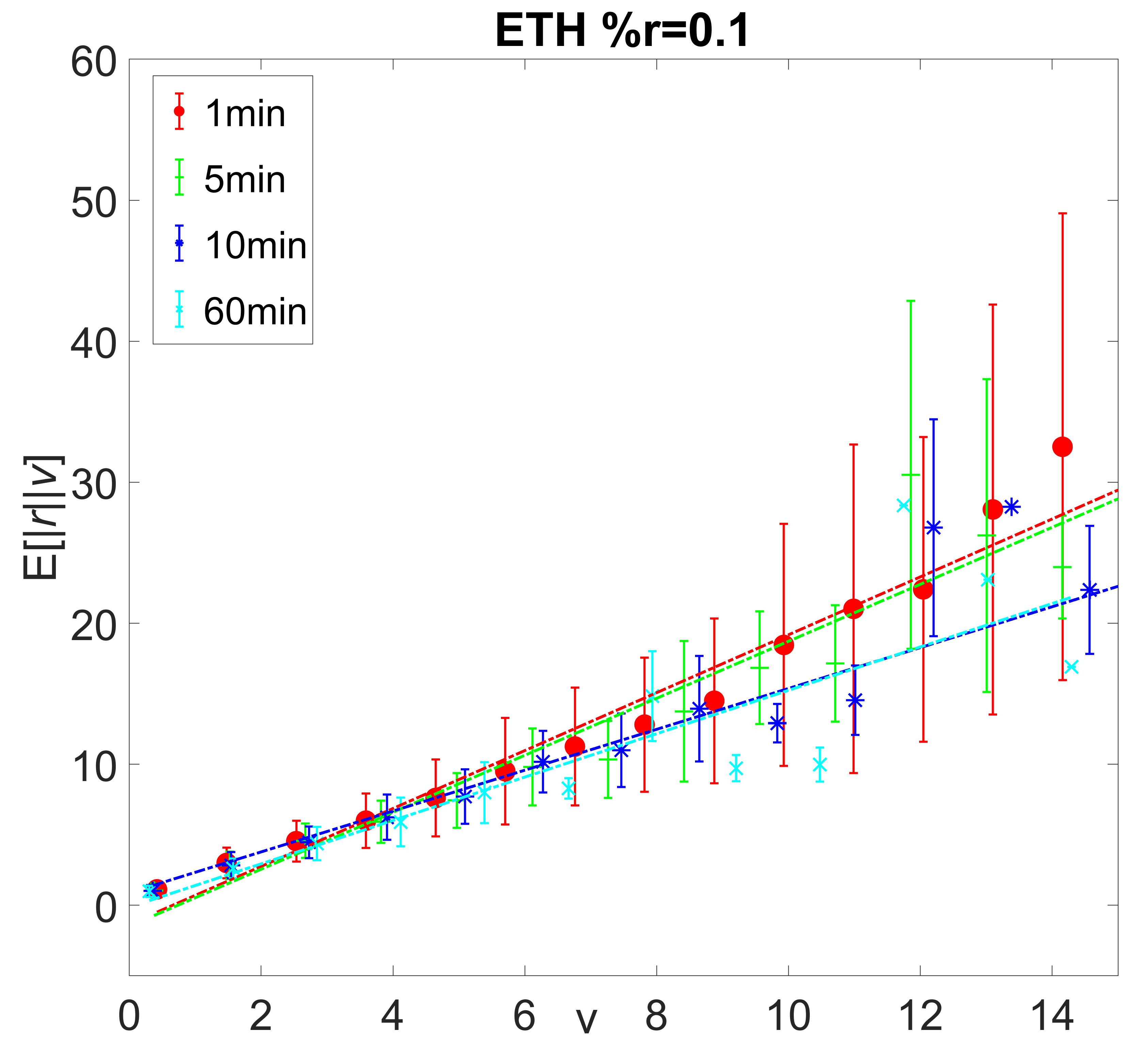}
\includegraphics[width=0.7\textwidth]{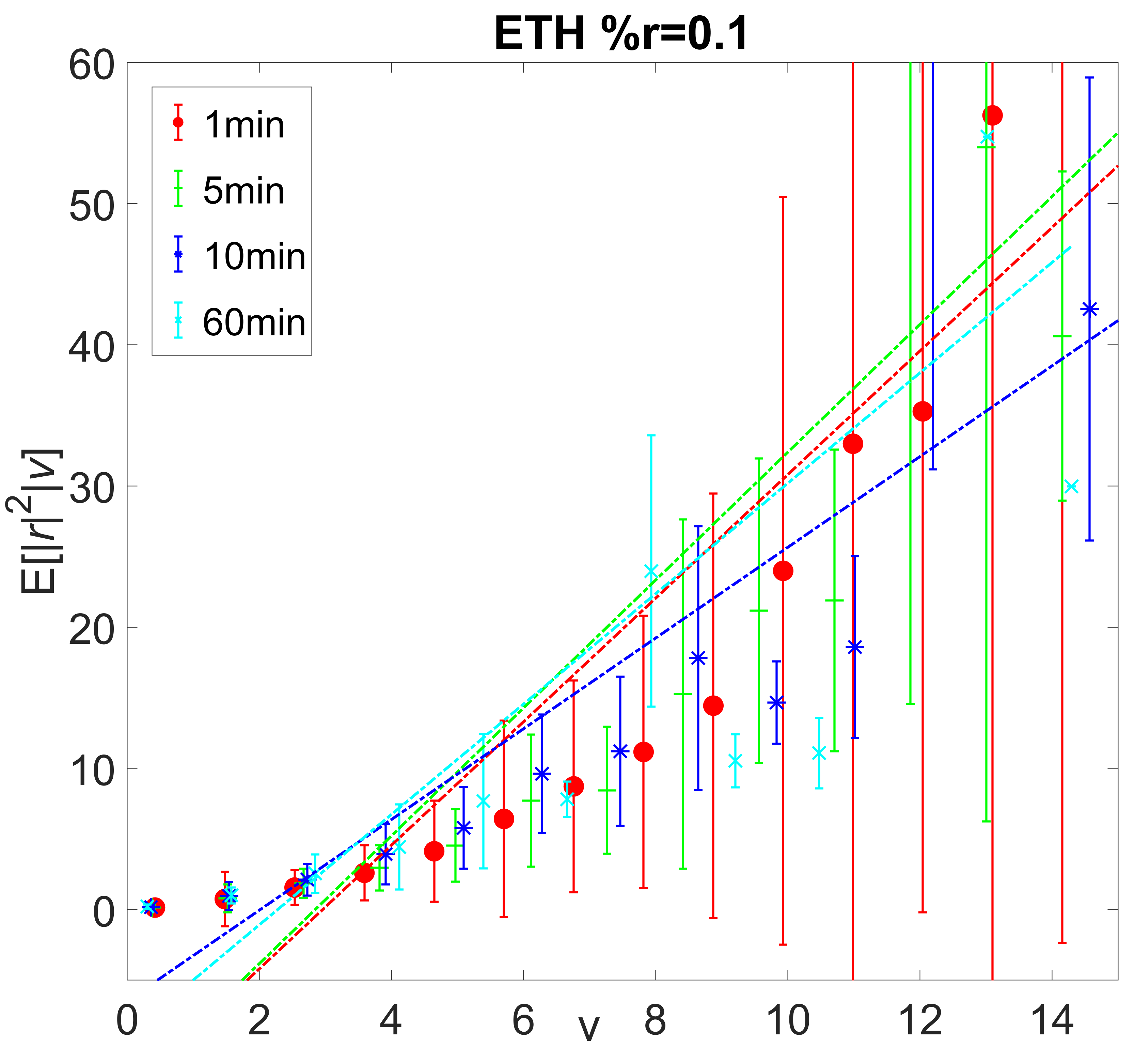}
\caption{The same quantities as in Fig.~\ref{fig::conditional.rv.BTC} for ETH.}
\label{fig::conditional.rv.ETH}
\end{adjustwidth}
\end{figure}

The values of the normalized volume traded $v_{\Delta t}(t)$ have been compartmentalized and in each cell $v_i$ a fixed fraction $p \ll 1$ of the respective largest conditional volatility values have been preserved for further analysis. A power-law function with the exponent $\kappa$ is assumed to model a relation between the two quantities:
\begin{equation}
v_{\Delta t} \sim |r_{\Delta t}|^{\kappa}, \quad |r_{\Delta t}| \sim v_{\Delta t}^{\alpha}.
\label{eq::price.impact.function}
\end{equation}
Fig.~\ref{fig::conditional.rv.BTC} tests whether any of the relations of the form $\mathbb{E}[|r_{\Delta t}|^{\kappa}|v_{\Delta t}] \sim v_{\Delta t}$ holds for BTC if the following exponent values were selected: $\kappa=0.2$, $\kappa=0.5$, $\kappa=1$, and $\kappa=2$. The threshold value has been chosen to be $p=0.1$, because for larger values the relation becomes blurred and difficult to identify, while for smaller values too few data points can be considered, which amplifies the uncertainty. Looking at the graphs, one can reject the hypothesis that volatility and volume are related via $v_{\Delta t} \sim |r_{\Delta t}|^2$ (i.e., $\alpha=0.5$) for all the sampling frequencies considered. In the case of the highest sampling frequency ($\Delta t=1$ min) the data is approximated the best for $\kappa=1$ and, secondarily, for $\kappa=0.5$ and $\kappa=0.2$, over the broad volume range: $1 < v_{\Delta t} < 16$. For $\Delta t \ge 10$ min none of the values considered of $\kappa$ works well, while for $\Delta t=5$ min two cases cannot be excluded: $\kappa=0.5$ and $\kappa=0.2$. This means that the likely functional form of the price impact cannot be inferred based on the available data. Fig.~\ref{fig::conditional.rv.ETH} presents the analogous results for ETH. The square-root form of the price impact (corresponding to $\kappa=2$) can also be rejected in this case. However, it cannot be decided which of the remaining models ($\kappa \le 1$) is the most likely.

The fact that $\kappa \ne 2$ ($\alpha \ne 0.5$) and likely $\kappa \le 1$ ($\alpha \ge 1$) for short sampling intervals is interesting, because it makes the price impact function linear or superlinear ($\alpha \ge 1$) -- a result that differs from some earlier claims made for the regular financial markets, where the function was concave, at least for short and moderate sampling intervals~\cite{Gabaix2003,BouchaudJP-2010a}. There is also a discrepancy for the long sampling intervals, because in this case the behaviour reported for the regular markets was effectively linear, while here it remains undefined. It is noteworthy in this context that the superlinear ($\alpha > 1$) price impact for large $\Delta t$ in Eq.~(\ref{eq::price.impact.function}) could open the space for market manipulation~\cite{BouchaudJP-2010a}, which on the cryptocurrency trading platforms can take the form of wash trading~\cite{KwapienJ-2022a,WashtradingPhysA}. According to that, one can view the presented results as being in favour of the conclusion that the full maturity is still ahead of the cryptocurrency market.

\subsection{Volatility clustering and long memory}

\begin{figure}[ht!]

\includegraphics[width=1\textwidth]{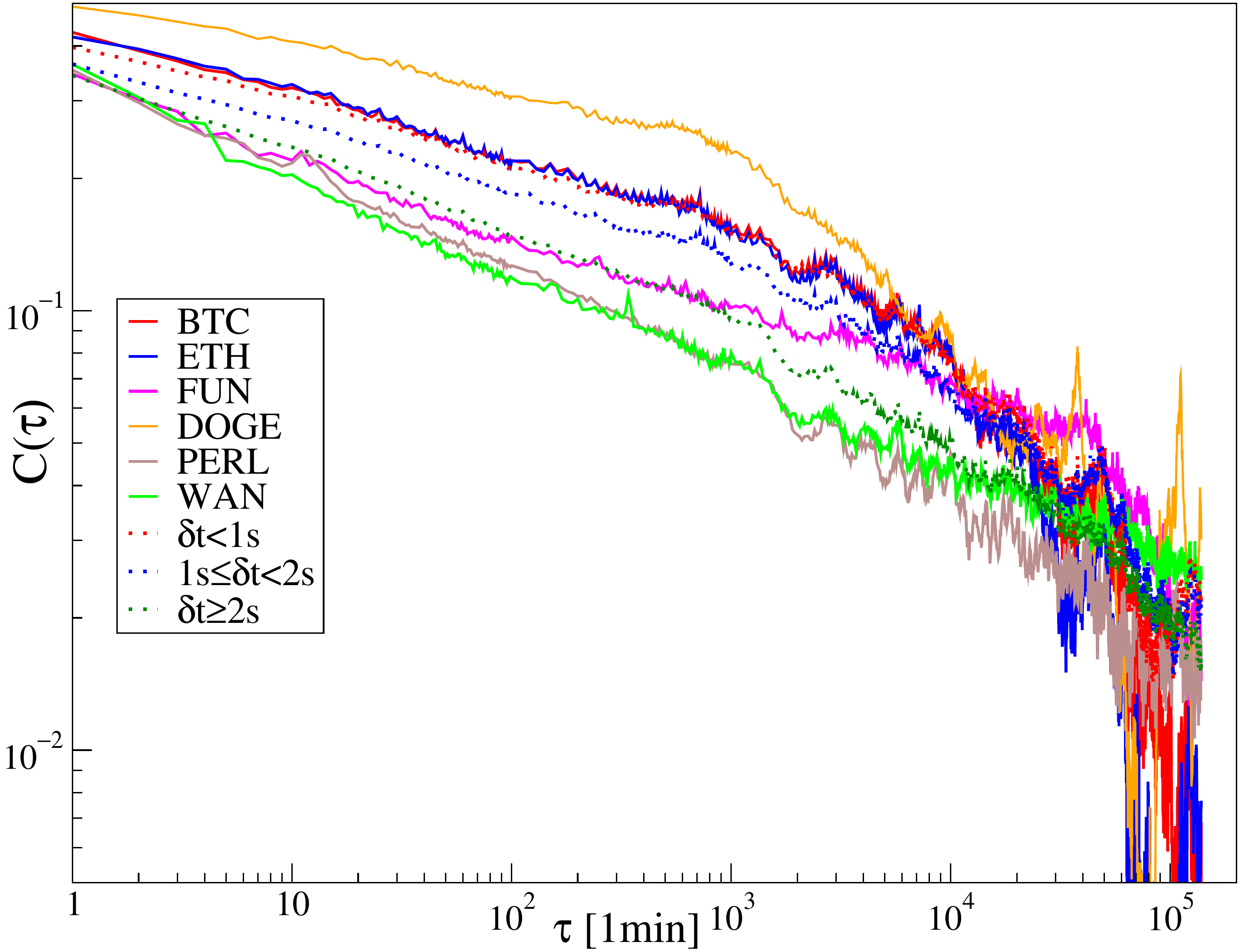}
\caption{Autocorrelation function $C_{|r_{\Delta t}|}(\tau)$ of the absolute normalized log-returns $|r_{\Delta t}(t)|$ (volatility) calculated for the selected individual cryptocurrencies---BTC, ETH, DOGE, FUN, PERL, and~WAN---as~well as for the cryptocurrency Groups I-III characterized by specific range of the average inter-transaction time: $\delta t<1s$ (Group I, dotted red), $1s \le \delta t<2s$ (Group II, dotted blue), $\delta t \ge 2s$ (Group III, dotted green). $C_{|r_{\Delta t}|}(\tau)$ has been averaged for each value of $\tau$ over all the cryptocurrencies belonging to a given group. Note the double-logarithmic~scale.}
\label{fig::autocorrelation.volatility}

\end{figure}

It takes some time for a market to absorb completely pieces of information that arrive there. This is a source of temporal market correlations that can be the easiest observed in the price fluctuation amplitudes. Correlations are responsible for the phenomenon of volatility clustering, i.e., the existence of prolonged periods of fluctuations with elevated amplitude that are separated by quiet periods with more tamed fluctuations~\cite{RakR-2007a}. Volatility clustering is observed on all markets and can quantified in terms of the autocorrelation function:
\begin{equation}
C(\tau) = \langle r_{\Delta t}(t) r_{\Delta t}(t-\tau) \rangle _t,
\label{eq::autocorrelation.function}
\end{equation}
where $\tau$ is the lag time. The autocorrelation functions calculated from the absolute log-returns for several individual cryptocurrencies and the average autocorrelation functions calculated for Groups I-III are presented in Fig.~\ref{fig::autocorrelation.volatility} on double logarithmic scale. In each case one can identify at least one range of lags over which $C(\tau)$ shows power-law decay. For BTC, ETH, and FUN there is only one such a range corresponding to $10 {\rm min} \le \tau \le 500$ min with a relatively small upper end, the same refers to WAN, but in this case the upper end exceeds $\tau \approx 20,000$ min (ca. two weeks). On the other hand, DOGE, PERL, and the average plots show two scaling regimes: the short-$\tau$ one up to $\tau \approx 500-1,000$ min (less than a day) and the long-$\tau$ one for $1,000 {\rm min} < \tau < 20,000$ min. In each case $C(\tau)$ falls to 0 around $\tau \approx 100,000$ min (more than 2 months). As compared to a more distant past, the scaling regions for BTC and ETH are shorter now (e.g., in the years 2016-2018 it reached $\tau=1,000$ min~\cite{DrozdzS-2019a}), which is consistent with the market time acceleration caused by increased trading frequency). This overall picture for the cryptocurrency market does not depart much from the one observed on other financial markets. A power-law decaying autocorrelation function expressing the long memory of volatility is a common property that is a manifestation of the way the market processes information~\cite{Gopikrishnan1999,DrozdzS-2009a}. The time lag at which $C(\tau)$ reaches a statistically insignificant level is equal to the average length of volatility cluster~\cite{DrozdzS-2009a}. Due to the alternating character of market dynamics, where the high-volatility periods are interwoven with the low-volatility periods, for larger time lags the autocorrelation function becomes negative. Note that due to the fact that volatility time series are unsigned, the long-range autocorrelations cannot be exploited for the related investment strategies.

\subsection{Multiscaling of returns}

\begin{figure}[ht!]
\includegraphics[width=1\textwidth]{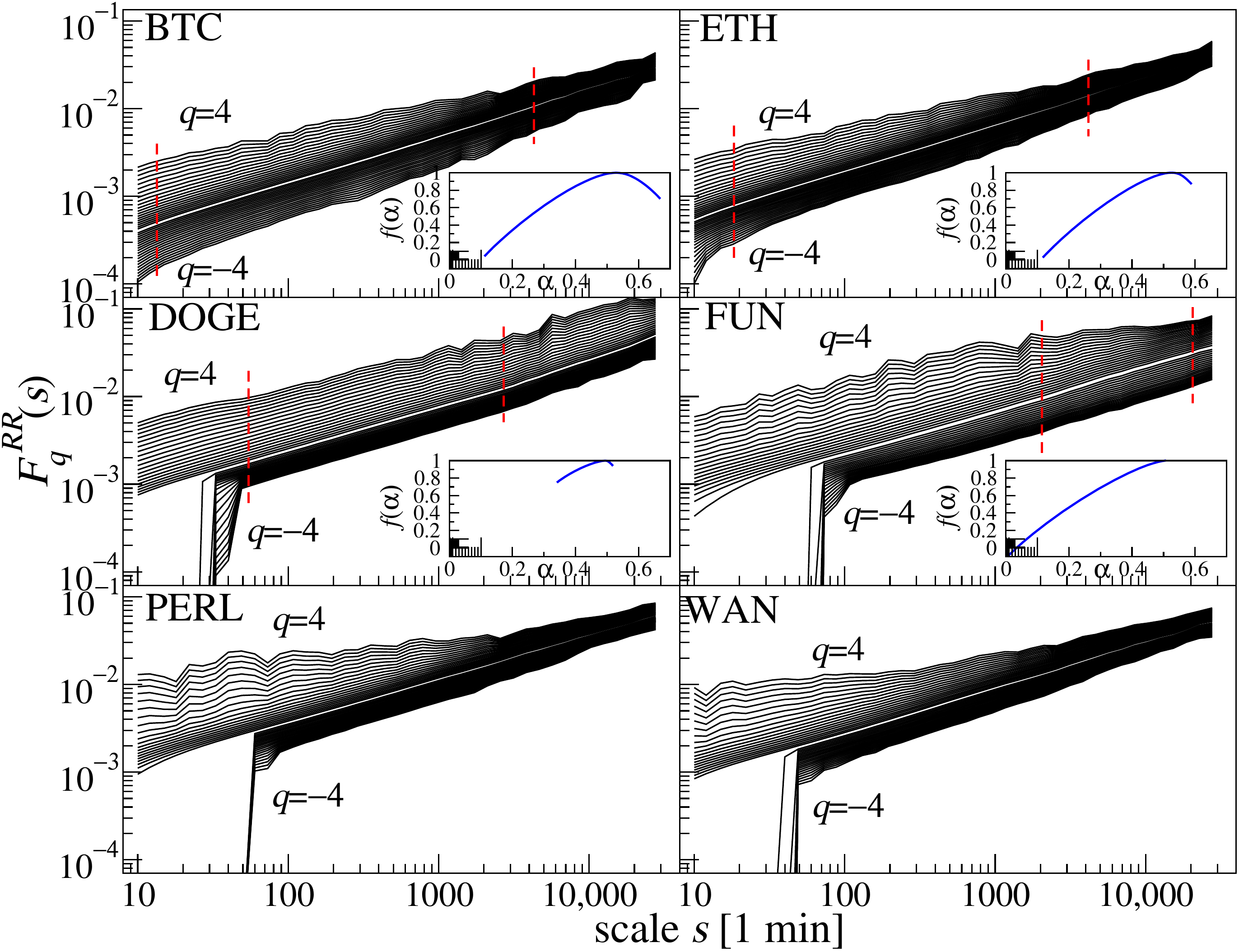}
\caption{(Main plots) Univariate fluctuation functions $F_q^{RR}(s)$ calculated from the log-returns $R_{\Delta t}(t)$ with $\Delta t=1$ min for BTC, ETH, DOGE, FUN, PERL, and~WAN. The~breakdown of scaling for small scales and negative values of $q$ in some plots is an artifact related to long sequences of zero returns in time series. (Insets) Singularity spectra $f(\alpha)$ calculated from the corresponding fluctuation functions in the range denoted by dashed red lines (if possible).}
\label{fig::fluctuation.functions}
\end{figure}

If the bivariate or univariate fluctuation functions can be approximated by a power-law relation
\begin{equation}
F_q^{AB}(s) \sim s^{h(q)},
\label{eq::fractal.scaling}
\end{equation}
where $h(q)$ is a non-increasing function of $q$ called the generalized Hurst exponent~\cite{kantelhardt2002} and $A$ and $B$ stand for either $R$ or $V$, the time series under study reveal fractal structure. If $h(q)={\rm const}=H$ it means that this structure is monofractal with $H$ equal to the Hurst exponent, which is a measure of persistence, otherwise it is multifractal~\cite{kantelhardt2002}. Multifractal signals are govern by the processes with long-range autocorrelations, that is why both properties are often observed together~\cite{Zhou2012,Klamut2018,Garcin2020,Kwapien2023}. It is so, for example, in the financial data. If the relation (\ref{eq::fractal.scaling}) exists, it can be seen in double logarithmic plots of $F_q^{AB}(s)$. Fig.~\ref{fig::fluctuation.functions} displays $F_q^{RR}(s)$ for six cryptocurrencies, with $-4 \le q \le 4$ and $10 \le s \le 25,000$. Out of these, 4 cryptocurrencies show unquestionable power-law functions: BTC, ETH, DOGE, and FUN for all used values of $q$ and for at least a decade-long range of scales, while PERL and WAN do not. The same result can be expressed in a different way by calculating the singularity spectra $f(\alpha)$ from $h(q)$ according to the following relations:
\begin{equation}
\alpha = h(q) + q {dh(q) \over dq}, \quad f(\alpha = q(\alpha - h(q)) + 1.
\label{eq::singularity spectra}
\end{equation}
The H\"older exponents $\alpha$ quantify the singularity strength and $f(\alpha_0)$ expresses the fractal dimension of a subset of singularities with strength $\alpha=\alpha_0$. While many theoretical singularity spectra are symmetric, in a practical situation, one observes spectra that are asymmetric~\cite{Kwapien2005,oswiecimka2006a,DrozdzS-2010a,oh2012multifractal,DrozdzBTC2018,Watorek2019,Jiang_2019}. The insets in Fig.~\ref{fig::fluctuation.functions} show $f(\alpha)$ calculated from $F_q^{RR}(s)$ in the scaling regions of $s$. All the presented spectra are left-side asymmetric (their left shoulder, corresponding to positive $q$s, is longer). The origin of such a behaviour can be twofold: the signals can develop heavy tails of the probability distribution functions that are unstable in the sense of L\'evy yet their convergence to the normal distribution is slow~\cite{DrozdzS-2009a} and the signals can be mixtures of the processes that have different fractal properties: large fluctuations can be associated with a multifractal process (e.g., a multiplicative cascade) while small fluctuations can be monofractal. It often happens that the small fluctuations of financial time series are noise while the medium and large fluctuations carry meaningful information.

It was reported in the literature that cryptocurrencies can also show such asymmetric $f(\alpha)$ spectra~\cite{DrozdzBTC2018,watorek2021}. From the perspective of this study, it is interesting to note that the spectra for BTC calculated for different historical periods show elongation of the right shoulder of $f(\alpha)$ that corresponds to small fluctuations. It can be interpreted as a gradual building of a multifractal structure in BTC price fluctuations that started from large returns only in the early stages of BTC trading and being imposed on the smaller ones as the cryptocurrency market goes towards maturity. If one looks at Fig.~\ref{fig::fluctuation.functions}, BTC and ETH, and to a lesser degree DOGE, that is, the cryptocurrencies that are among the most capitalized ones, have noticeable right wings of $f(\alpha)$, while the more exotic cryptocurrencies like FUN, PERL, and WAN do not develop the right wing at all. In agreement with what has been said before, despite that various cryptoassets are traded on the same platforms, the different ones can be found at different stages of the maturing process due to the different trading frequency. This difference can also be observed in the possible scaling range of the fluctuation functions in Fig.~\ref{fig::fluctuation.functions}. In the case of the two most liquid cryptocurrencies, BTC and ETH, the $F_q^{RR}(s)$ scaling can be observed almost from the beginning of the scale range, whereas in the case of less liquid cryptocurrencies the range of satisfactory scaling is significantly shorter and $F_q^{RR}(s)$ even become singular on short scales due to the number of consecutive 1-min bins with zero returns. This is typical behavior in the case of less liquid financial instruments~\cite{DrozdzBTC2018}.

\subsection{Cross-correlations among cryptocurrencies}

Information that flows into the market may have the same impact on certain assets that, for example, share similar characteristics, such as the market sector, the main shareholders, or, in the case of cryptocurrency, the type or consensus mechanism~\cite{JAMES2022PhysA}. This can lead to the emergence of cross-correlation between such assets and to a certain hierarchy of cross-correlations (e.g., sector, subsector, and bilateral ones)~\cite{kwapien2012}. The correlation structure is a dynamical property that can change suddenly and substantially over time as the market reacts to perturbations~\cite{James2022PhysicaD}. In quiet times it is well-shaped, elastic and hierarchical, while during periods of turmoil, it becomes centralized and rigid. This dual behaviour is characteristic for the developed markets, while a lack of cross-correlations or a persistent centralization may be attributed to immaturity.

\begin{figure}[ht!]
\begin{adjustwidth}{-\extralength}{0cm}

\includegraphics[width=1.35\textwidth]{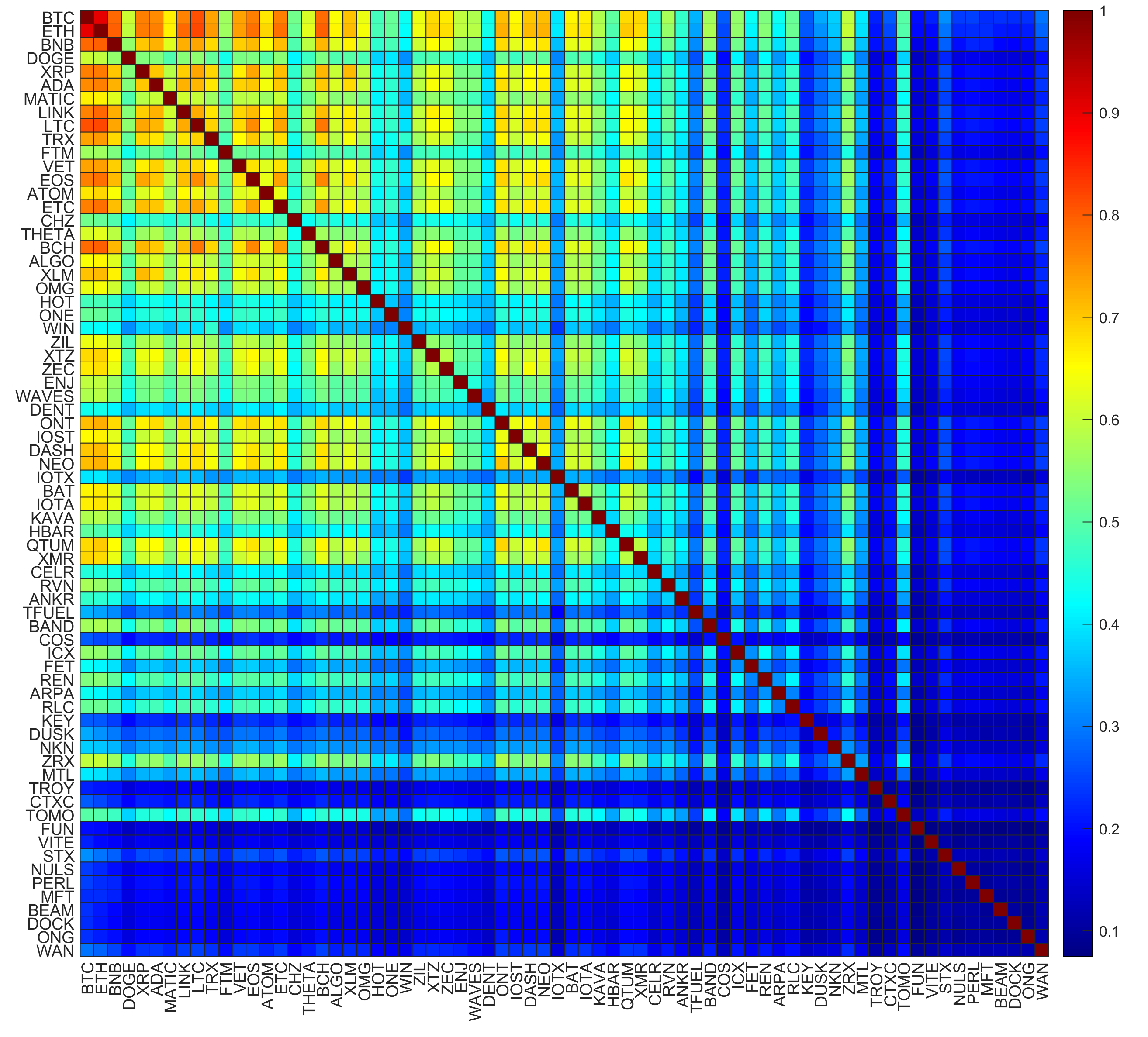}
\end{adjustwidth}
\caption{The $q$-dependent detrended cross-correlation matrix entries $\rho_q^{ij}(s)$ calculated from time series of log-returns representing 70 cryptocurrencies with $q=1$ and $s=10$ min. Cryptocurrencies have been sorted according to the average inter-transaction time $\delta t$ in increasing order (top to bottom). The~color-coding scheme is shown on the~right.}
\label{fig::rhoq.matrix.entries.crypto}

\end{figure}

The market cross-correlation structure can concisely be characterised by the matrix approach. For a set of $N$ time series of log-returns representing different cryptocurrencies, $N(N-1)/2$ the $q$-dependent detrended cross-correlation coefficients $\rho_q^{ij}(s)$ can be calculated, where $i,j=1,...,N$ and $\rho_q^{ij}=\rho_q^{ji}$, which form a $q$-dependent detrended cross-correlation matrix ${\bf C}_q(s)$. Due to the fact that the cross-correlation strength increases typically with scale for all the asset pairs, the differences in $\rho_q^{ij}(s)$ are on average minimal for the shortest studied scale of $s=10$ min. However, even in this case, it is easy to observe that different cryptocurrency pairs reveal strong differences. Fig.~\ref{fig::rhoq.matrix.entries.crypto} presents the complete matrix ${\bf C}_q(s)$ with the cryptocurrencies ordered according to the average inter-transaction time $\langle \delta t \rangle_t$. The ordering allows one to find even by eye a significant cross-correlation between $\langle \delta t \rangle_t$ and $\rho_q^{ij}$: the shorter this time is, the stronger the cross-correlations are. In full analogy to other markets, information needs time to propagate over the whole cryptocurrency market and the propagation speed is crucially dependent on the cryptocurrency liquidity that can roughly be approximated by the transaction number per time unit. Based on the exact values of $\rho_q^{ij}(s)$, one can notice that even the least frequently traded cryptocurrencies from the considered basket develop statistically significant dependencies among themselves. This, however, might not be true for even less capitalized tokens, which can have an idiosyncratic dynamics.

\begin{figure}[ht!]
\begin{adjustwidth}{-\extralength}{0cm}

\includegraphics[width=0.67\textwidth]{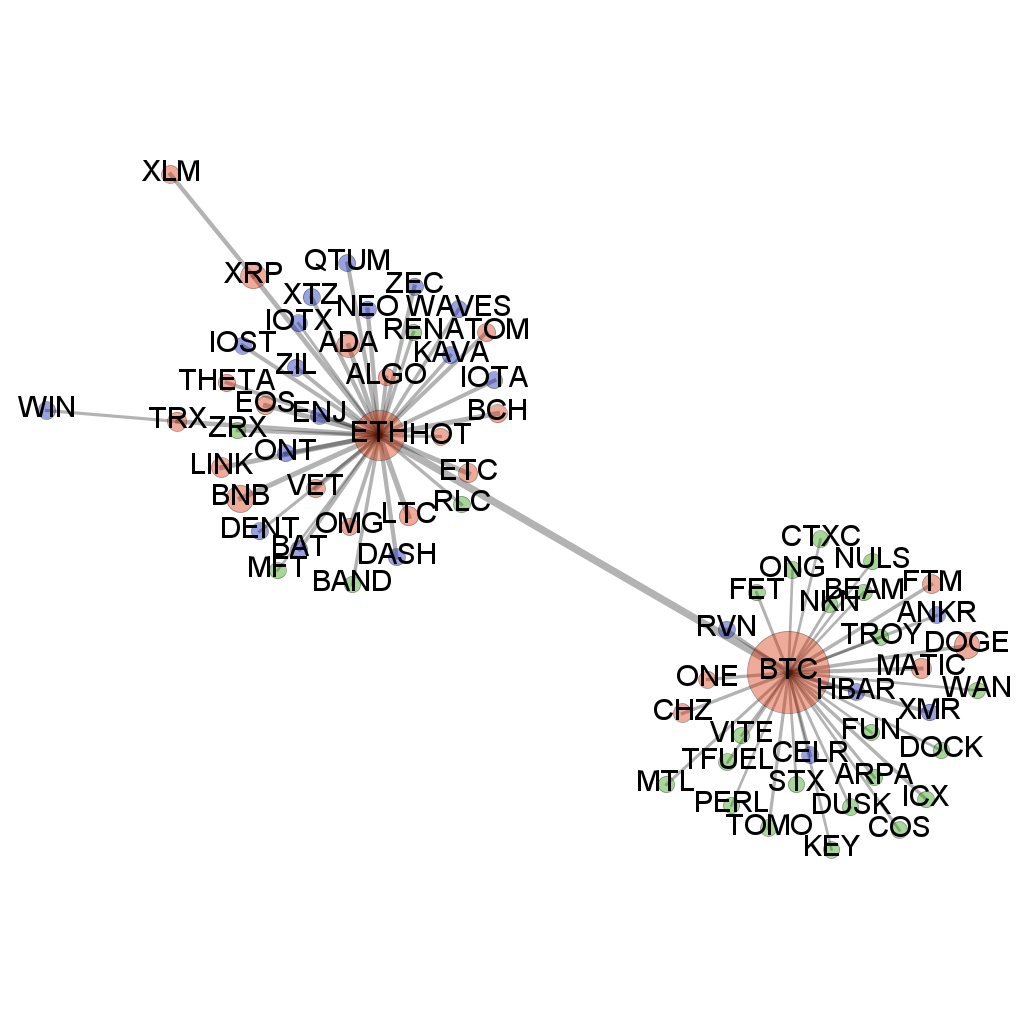}
\includegraphics[width=0.67\textwidth]{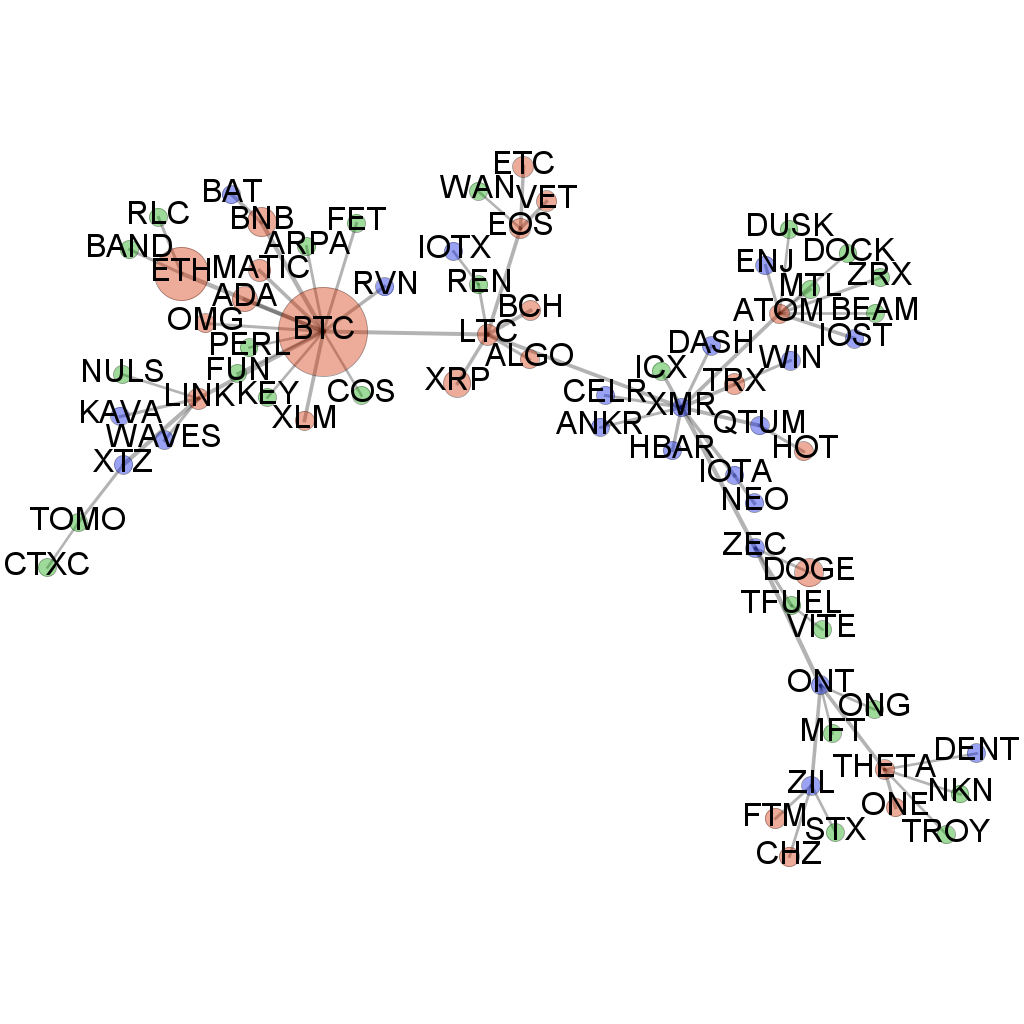}
\end{adjustwidth}
\caption{Minimal spanning trees calculated from a distance matrix ${\bf D}_q(s)$ based on $\rho_q(s)$ for $s=10$ and for $q=1$ (left) and $q=4$ (right). Within~each tree, the~size of the vertex is proportional to the average value of the volume $W_{\Delta t}$ for $\Delta t=1$ min, while the width of the edge is proportional to $1-d_q^{ij}(s)$. The~vertex sizes cannot be directly compared across the trees, however. Colors represent Groups I-III in terms of the trading frequency: $\delta t<1s$ (Group I, red), $1s\le \delta t<2s$ (Group II, blue), and~$\delta t \ge 2s$ (Group III, green).}
\label{fig::mst.crypto}

\end{figure}

The correlation matrix ${\bf C}_q(s)$ can be transformed into a distance matrix ${\bf D}_q(s)$ with the entries
\begin{equation}
d^{ij}_q(s)=\sqrt{2(1-\rho^{ij}_q(s))},
\label{eq::metric-distance}
\end{equation}
which differs from the former in that its entries $d_q^{(ij)}$ are metric. ${\bf D}_q(s)$ can be used for constructing a weighted graph with nodes representing cryptocurrencies and edges representing distances $d_q^{(ij)}(s)$. Next, by using the Prim algorithm~\cite{prim1957} one can construct the corresponding $q$-dependent detrended minimal spanning tree (MST), which can be considered as a connected minimum-weight subset of the graph containing all $N$ nodes and $N-1$ edges. The MST topology depends strongly on the cross-correlation structure of a market. A centralized market corresponds to a star-like MST, while a market containing idiosyncratic assets shows an MST with elongated branches and no dominant hubs. Fig.~\ref{fig::mst.crypto} exhibits two MSTs created from all the 70 cryptocurrencies for $q=1$ (left) and $q=4$ (right). The former involves cross-correlations between the fluctuations of all magnitudes, while the latter involves only the large fluctuations. For $q=1$ the structure is dual star with BTC and ETH as its central hubs. there is no surprise here as both cryptocurrencies are distinguished by their fame and large capitalization, which makes them a kind of reference for the remaining cryptocurrencies. On the other hand, for $q=4$ the structure is more distributed with a primary hub, BTC, and a few secondary ones: LTC, XMR, and ONT. This means th4at the relatively large fluctuations are not collectively correlated, unlike the majority of fluctuations, and more subtle dependencies are present. This goes in parallel with the conclusions based on the multifractal analysis, where those were the large fluctuations that carried the clearly multifractal characteristics and long-term correlations, while the small fluctuations were much more noisy. It is worth to mention that a similar behaviour can be observed in the stock market, where the cross-correlation structure carried by the large fluctuations is much richer than that carried by the medium and small fluctuations~\cite{kwapien2017}. However, in the stock market, the heterogeneous cross-correlation structure is more pronounced even in the latter case~\cite{kwapien2017,JAMES2022PhysA}. Since there is no clear division into market sectors~\cite{Chaudhari2020}, the cryptocurrency market appears to be less developed from this particular perspective.

\subsection{Cross-correlations between cryptocurrencies and the other markets}

Recently, BTC and ETH have been found to be significantly coupled to the traditional financial markets during the period covering the Covid-19 pandemic and the bear market of 2022~\cite{WatorekM-2023a}. This result has essential practical implications in risk management as cryptocurrencies cannot serve as hedging assets~\cite{James2023}. It differs from earlier findings that before 2020 the cryptocurrency market was detached from the traditional markets~\cite{Corbet2018,urquhart2019,manavi2020,DrozdzS-2020a}, but at the same time it remains in agreement with the observations that Covid-19 changed the safe-haven paradigm and contributed to the correlation of major cryptocurrencies with traditional risk assets~\cite{Kristoufek2020,James2021,Yarovaya2022,Wang2022,Zitis2023}. So far only the most capitalized cryptocurrencies were studied~\cite{WatorekM-2023a} and this is why here the cryptocurrencies with smaller capitalization are also studied.

\begin{figure}[ht!]
\begin{adjustwidth}{-\extralength}{0cm}

\includegraphics[width=1.32\textwidth]{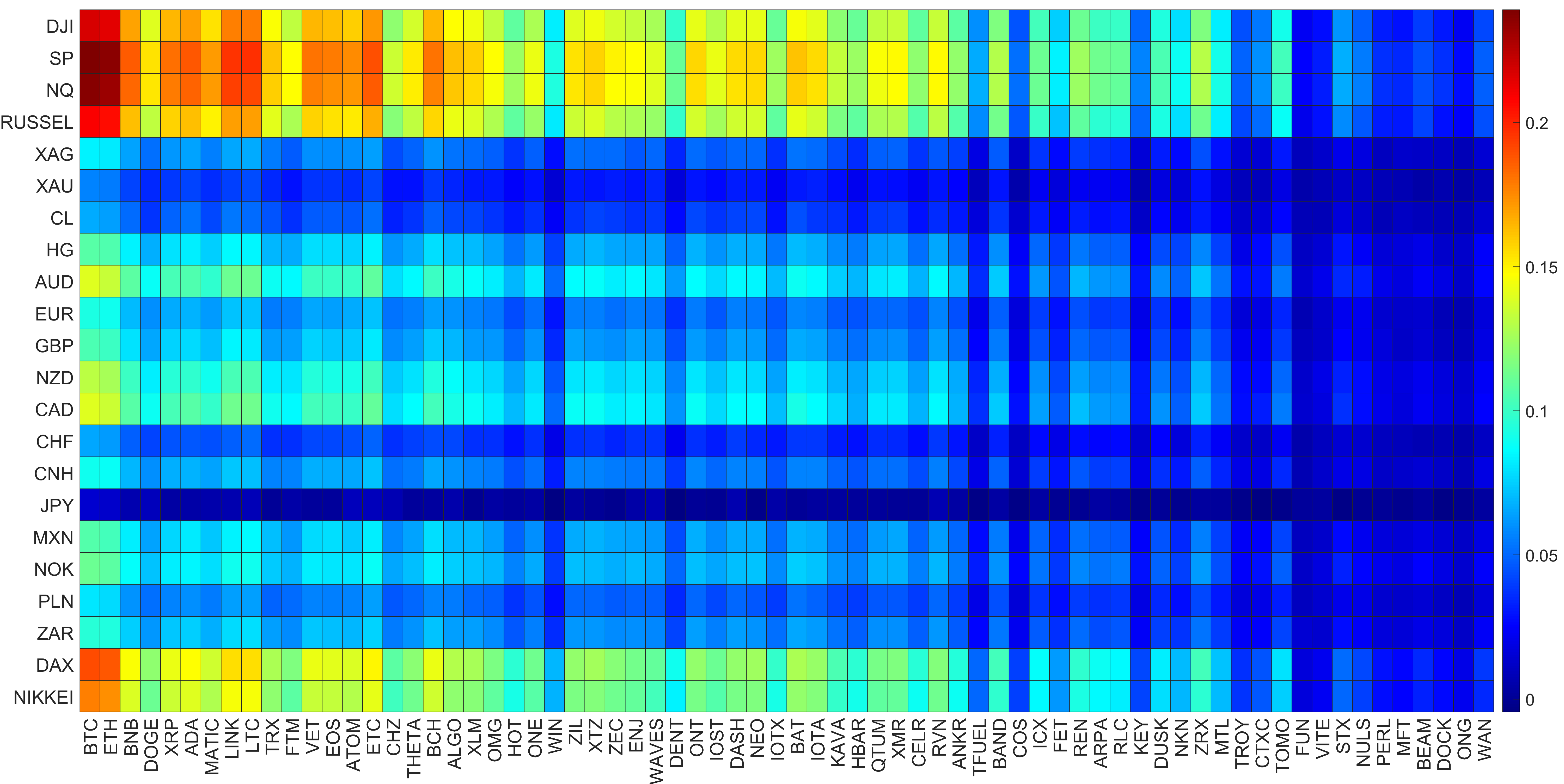}
\end{adjustwidth}
\caption{The $q$-dependent detrended cross-correlation matrix entries $\rho_q^{ij}(s)$ calculated from time series of log-returns representing selected cryptocurrencies and selected traditional financial instruments with $q=1$ and $s=10$ min. Cryptocurrencies have been sorted according to the average inter-transaction time $\delta t$ in increasing order (top to bottom). The~color coding scheme, which differs from the one in Figure~\ref{fig::rhoq.matrix.entries.crypto}, is shown on the~right.}
\label{fig::rhoq.matrix.entries.regular}

\end{figure}

Time series of log-returns of 70 cryptocurrencies and 22 traditional financial instruments have been collected from Dukascopy platform~\cite{Dukascopy}. Among the latter, there are contracts for differnce (CFDs) representing the returns of 12 fiat currencies (AUD, CAD, CHF, CNH, EUR, GBP, JPY, MXN, NOK, NZD, PLN, and ZAR), 4 commodities - WTI crude oil (CL), high grade copper (HG), silver (XAG), and gold (XAU), 4 US stock market indices - Nasdaq 100 (NQ100), S\&P500, Down Jones Industrial Average (DJI), and Russell 2000 (RUSSEL), the main German stock index DAX 40 (DAX), and the Japanese Nikkei 225 (NIKKEI). All these instruments except for the non-US stock indices have been expressed in USD. Their quotes cover a period from Jan 1, 2020 to Dec 30, 2022. The quotes were recorded over the trading hours, i.e., from Sunday 22:00 to Friday 20:15 UTC with a break between 20:15 and 22:00 UTC each trading day. To assess the cross-correlations, the cryptocurrency time series were synchronized with the those from Dukascopy. Cross-correlations have been quantified by $\rho_q^{RR}(s)$. 

\begin{figure}[ht!]
\includegraphics[width=0.95\textwidth]{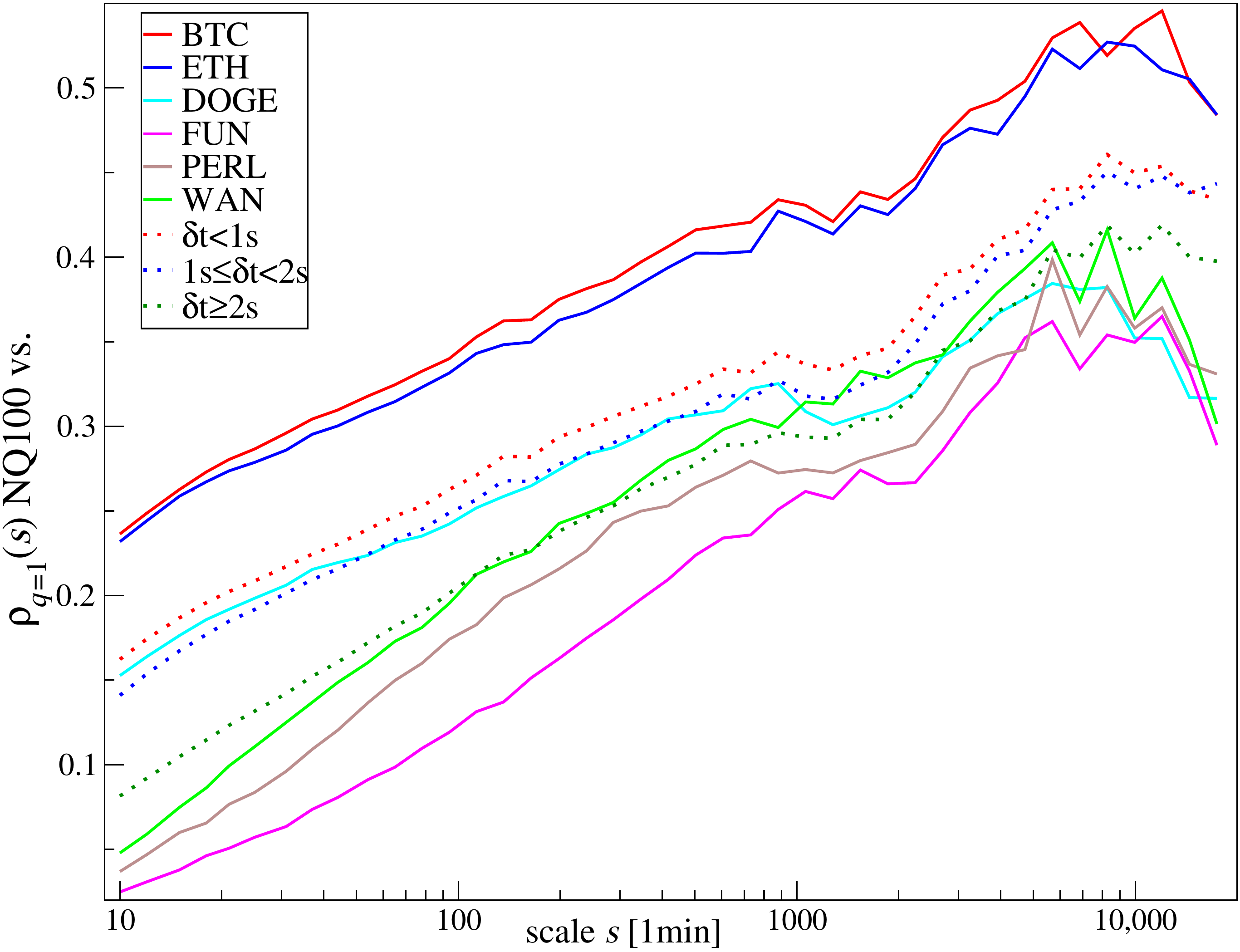}
\caption{The $q$-dependent detrended cross-correlation coefficient $\rho_q^{RR}(s)$ calculated for the pairs of log-return time series consisting of NASDAQ 100 and a cryptocurrency (BTC, ETH, DOGE, FUN, PERL, or~WAN) or a group of cryptocurrencies characterized by average inter-transaction time from a specific range: $\delta t<1s$ (Group I, red), $1s\le \delta t<2s$ (Group II, blue), and~$\delta t \ge 2s$ (Group III, green).}
\label{fig::rho.q.rr.nasdaq}
\end{figure}

Fig.~\ref{fig::rhoq.matrix.entries.regular} shows the $q$-dependent detrended cross-correlation matrix ${\bf C}_q(s)$ entries for the inter-market pairs consisting of a cryptocurrency and a traditional asset. The first observation is that the maximum available values of the matrix entries do not exceed $\rho_q^{RR}(s)=0.25$, which makes them much smaller that in the case of the inner cross-correlation among the cryptocurrencies. This is an expected effect, because markets are typically more tightly coupled inside than outside. Among the strongest cross-correlations, one can point out the coupling of BTC and ETH with the American stock market indices ($\rho_q^{RR}(s) > 0.2$ and with NIKKEI and DAX ($0.15 < \rho_q^{RR}(s) < 0.2$). Considerably weaker yet still prominent are the cross-correlations between several other cryptocurrencies like XRP, ADA, LTC, LINK, VET, ETC, EOS, ATOM, and BCH on one side and the American indices ($0.15 < \rho_q^{RR}(s) < 0.2$). The relations between cryptocurrencies and fiat currencies remain moderate with the AUD, CAD, and NZD being relatively the strongest ($0.1 < \rho_q^{RR}(s) < 0.15$). Contrary to it, the cryptocurrencies are the most decoupled from JPY, CHF, gold (XAU), and crude oil (CL). A general observation is that the less liquid a cryptocurrency is, the weakest is its cross-correlation with the traditional instruments. Here again, DOGE is somewhat of an exception and has a weaker cross-correlation than its trading frequency and capitalization would imply. However, it should be noted that the values collected in Fig.~\ref{fig::rhoq.matrix.entries.regular} correspond to the shortest available scale of $s=10$ min. How these values refer to the maximum cross-correlations for longer scales Fig.~\ref{fig::rho.q.rr.nasdaq} documents. Here the cross-correlation between the selected cryptocurrencies and their sets grouped according to the average inter-transaction time (Groups I-III) and NASDAQ 100 is presented. This particular choice of the traditional index has been motivated by the fact that the cryptocurrency market is strongly cross-correlated with it~\cite{WatorekM-2023a}. Indeed, for much longer $s$ the values of $\rho_q^{RR}(s)$ grow significantly and even reach some saturation level resembling the Epps effect for $s > 500$ min with the average values of $\rho_q^{RR}(s)$ in Groups I-III oscillating around 0.4 (for a given scale, $\rho_q^{RR}(s)$ decreases systematically with increasing $\delta t$). The cryptocurrencies that are the most cross-correlated with NASDAQ 100, i.e., BTC and ETH, have the maximum values of $\rho_q^{RR}(s) > 0.5$.

\section{Conclusions}

The statistical properties of price log-returns and volume of the cryptocurrencies were the central point of the present study. The existence of the so-called financial stylized facts in the cryptocurrency market during the last 3 years was investigated and compared with the stylized facts observed in the traditional financial markets. Several characteristics were of particular interest: a tail behaviour of the probability distribution functions for the log-returns and volume traded, the functional form of price impact, volatility autocorrelations, multiscaling, cross-correlations among the cryptocurrencies, and cross-correlations between the principal cryptocurrencies and selected traditional market assets. Almost all the analyzed characteristics of the cryptocurrency market were found to be in qualitative agreement with their counterparts from the traditional markets. It allows one to conclude that, from this particular perspective, the cryptocurrency market does not differ from the mature markets.

Despite such a positive conclusion, one still has to be cautious. First, the level of the maturity of the cryptocurrencies depends on their trading frequency. The most liquid ones, such as BTC and ETH, to a greater extent, have characteristics corresponding to mature financial markets, and the least liquid ones do not. Second, the price impact function, while also of a power-law form, occurs to be substantially different from its counterparts reported in the traditional markets (linear or convex here vs. concave there~\cite{BouchaudJP-2010a}). Third, while the statistical properties are important from a practical point of view, as they can be exploited in various investment strategies, there are nevertheless many other important indicators of market maturity that were not investigated here. For example, the number of cryptocurrencies traded on the largest platforms, like Binance, is so large that it already matches the world's largest markets like New York Stock Exchange and NASDAQ. On the other hand, even the most recognized cryptocurrencies, like BTC and ETH, show extreme volatility, which means that the market is still rather illiquid and this property can question its maturity. There is another problem associated with the fact that the cryptocurrencies are often viewed as the speculation toys rather than the full-scale investment instruments. There are also numerous issues related to the limited reliability of the cryptocurrencies, their weak supply elasticity, etc. These problems, while important, were beyond the scope of this analysis, which one has to keep in mind when thinking about the given conclusions. Repeating this kind of analysis in future in order to follow how the cryptocurrency market changes seems to be a straightforward direction of potential future studies.

\vspace{6pt} 

\authorcontributions{Individual contributions of the authors are as follows. Conceptualization, S.D. and M.W.; methodology, S.D., J.K., and M.W.; software, M.W.; validation, S.D., J.K., and M.W.; formal analysis, S.D. and M.W.; investigation, S.D., J.K., and M.W.; resources, M.W.; data curation, M.W.; writing---original draft preparation, J.K.; writing---review and editing, S.D., J.K., and M.W.; visualization, M.W.; supervision, S.D. and J.K.; project administration, S.D. and M.W. All authors have read and agreed to the published version of the manuscript.}

\funding{This research received no external funding.}

\dataavailability{The data used in the article are freely available on Binance and Dukascopy platforms.} 

\conflictsofinterest{The authors declare no conflict of interest.}

\newpage
\appendixtitles{no}
\appendixstart
\appendix
\section[\appendixname~\thesection]{}
\label{sect::appendix}

\begin{table}[ht!]
\caption{List of cryptocurrencies from Binance.}
\begin{tabular}{|l|l||l|l||l|l|}
\hline
Ticker          & Name                 & Ticker          & Name          & Ticker        & Name        \\ \hline\hline
ADA             & cardano              & FET             & fetch         & QTUM          & qtum        \\ \hline
ALGO            & algorand             & FTM             & fantom        & REN           & ren         \\ \hline
ANKR            & ankr                 & FUN             & funtoken      & RLC           & iexec       \\ \hline
ARPA            & arpa chain           & HBAR            & hedera        & RVN           & ravencoin   \\ \hline
ATOM            & cosmos               & HOT             & holo          & STX           & stacks      \\ \hline
BAND            & band protocol        & ICX             & icon          & TFUEL         & theta fuel  \\ \hline
BAT             & basic atention token & IOST            & iost          & THETA         & theta       \\ \hline
BCH             & bitcoin cash         & IOTA            & miota         & TOMO          & tomochain   \\ \hline
BEAM            & beam                 & IOTX            & iotex         & TROY          & troy        \\ \hline
BNB             & binance coin         & KAVA            & kava          & TRX           & tron        \\ \hline
BTC             & bitcoin              & KEY             & key           & VET           & vechain     \\ \hline
CELR            & celer network        & LINK            & chainlink     & VITE          & vite        \\ \hline
CHZ             & chiliz               & LTC             & litecoin      & WAN           & wanchain    \\ \hline
COS             & contentos            & MATIC           & polygon       & WAVES         & waves       \\ \hline
CTXC            & cortex               & MFT             & hifi finance  & WIN           & winklink    \\ \hline
DASH            & dash                 & MTL             & metal         & XLM           & stellar     \\ \hline
DENT            & dent                 & NEO             & neo           & XMR           & monero      \\ \hline
DOCK            & dock                 & NKN             & nkn           & XRP           & ripple      \\ \hline
DOGE            & dogecoin             & NULS            & nuls          & XTZ           & tezos       \\ \hline
DUSK            & dusk network         & OMG             & omg network   & ZEC           & zcash       \\ \hline
ENJ             & enj coin             & ONE             & harmony       & ZIL           & zilliqa     \\ \hline
EOS             & eos                  & ONG             & ontology gas  & ZRX           & 0x          \\ \hline
ETC             & ethereum classic     & ONT             & ontology      &               &             \\ \hline
ETH             & ethereum             & PERL            & perl          &               &             \\ \hline
\end{tabular}
\label{tab::ticker.list}
\end{table}

\begin{adjustwidth}{-\extralength}{0cm}

\reftitle{References}

\bibliography{refs}

\end{adjustwidth}

\end{document}